\documentclass[acmsmall]{acmart}

\AtBeginDocument{%
  \providecommand\BibTeX{{%
    \normalfont B\kern-0.5em{\scshape i\kern-0.25em b}\kern-0.8em\TeX}}}

\setcopyright{cc}
\setcctype{by-nc-nd}
\acmJournal{PACMHCI}
\acmYear{2026} \acmVolume{10} \acmNumber{2} \acmArticle{CSCW014} \acmMonth{4} \acmPrice{}\acmDOI{10.1145/3788050}

\usepackage{CJKutf8}
\usepackage{array}
\usepackage{tablefootnote}
\newcolumntype{L}[1]{>{\raggedright\let\newline\\\arraybackslash\hspace{0pt}}m{#1}}
\newcolumntype{C}[1]{>{\centering\let\newline\\\arraybackslash\hspace{0pt}}m{#1}}
\newcolumntype{R}[1]{>{\raggedleft\let\newline\\\arraybackslash\hspace{0pt}}m{#1}}
\renewcommand{\arraystretch}{1.1}
\usepackage{colortbl} 
\usepackage{adjustbox}
\usepackage{multirow}
\usepackage{array, xcolor, longtable}
\renewcommand{\arraystretch}{1.1} 

\definecolor{mypink2}{RGB}{239, 240, 255}
\begin{document}

\title[Dissolving a Digital Relationship]{Dissolving a Digital Relationship: A Critical Examination of Digital Severance Behaviours in Close Relationships}

\author{Michael Yin}
\authornote{Both authors contributed equally to this research.}
\affiliation{
  \institution{University of British Columbia}
  \city{Vancouver}
  \state{BC}
  \country{Canada}
}
\email{jiyin@cs.ubc.ca}
\orcid{0000-0003-1164-5229}

\author{Angela Chiang}
\authornotemark[1]
\affiliation{
  \institution{University of British Columbia}
  \city{Vancouver}
  \state{BC}
  \country{Canada}
}
\email{achi2048@student.ubc.ca}
\orcid{0009-0008-8960-5837}

\author{Robert Xiao}
\affiliation{
  \institution{University of British Columbia}
  \city{Vancouver}
  \state{BC}
  \country{Canada} 
}
\email{brx@cs.ubc.ca}
\orcid{0000-0003-4306-8825}

\begin{abstract}
Fulfilling social connections are crucial for human well-being and belonging, but not all relationships last forever. As interactions increasingly move online, the act of digitally severing a relationship --- e.g. through blocking or unfriending --- has become progressively more common as well. This study considers actions of ``digital severance'' through interviews with 30 participants with experience as the initiator and/or recipient of such situations. Through a critical interpretative lens, we explore how people perceive and interpret their severance experience and how the online setting of social media shapes these dynamics. We develop themes that position digital severance as being intertwined with power and control, and we highlight (im)balances between an individual's desires that can lead to feelings of disempowerment and ambiguous loss for both parties. We discuss the implications of our research, outlining three key tensions and four open questions regarding digital relationships, meaning-making, and design outcomes for future exploration.
\end{abstract}




\maketitle

\section{Introduction}

Fulfilling relationships are important for human well-being. Such relationships provide personal support \cite{apostolouWhyPeopleMake2021, lynchQUALITATIVESTUDYQUALITY2008, overallHelpingEachOther2010}, improve emotional and physical health \cite{clearyFriendshipMentalHealth2018, holt-lunstadWhySocialRelationships2018, siasFriendshipSocialSupport2007, gomez-lopezWellBeingRomanticRelationships2019}, and drive psychological fulfillment \cite{carberyFriendshipNeedFulfillment1998, demirFriendshipNeedSatisfaction2010, kyleEnduringLeisureInvolvement2004, leNeedFulfillmentEmotional2001}. However, relationships also require time and maintenance \cite{roseKeepingEndingCasual1986}. Becoming so invested in a relationship can blind one to another's faults, leading to negative effects on well-being \cite{healyWeReJust2015}. Even the closest relationships can fracture, and the end of a relationship can be marked by sudden conflict, repeated strife, or simply drifting \cite{healyWeReJust2015, roseKeepingEndingCasual1986, roseHowFriendshipsEnd1984, apostolouThisHasEnd2023}. As close relationships dissolve, individuals hold conflicting emotions such as sadness and distress, sometimes mixed with happiness and relief \cite{flanneryBreakingFriendHard2021, bowkerWhenBestFriendships2024, bowkerExploratoryStudyBest2023}. 

As relationships have increasingly moved online, researchers have explored the effects of the online setting on the authenticity \cite{haimsonOnlineAuthenticityParadox2021} and depth \cite{chanComparisonOfflineOnline2004, tangDevelopmentOnlineFriendship2010} of interactions. Social media transforms dynamics and expectations regarding the permanence and significance of connections \cite{nesiTransformationAdolescentPeer2018, walther2011theories}, flattening multidimensional real-world connections into discrete digital labels such as friends or followers \cite{hoganBreakUpsLimitsEncoding2017}. These labels are easily established with a single click, creating a status indicator of an online connection and a medium for communication and updates \cite{gashi2016unfriending}. While this bi-directional contract forms the initial digital connection \cite{gashi2016unfriending}, a one-sided severance marks the end of this connection. As social exclusion (or \emph{ostracism}) has negative effects \cite{williamsOstracismConsequencesCoping2011a, williamsOstracismSocialExclusion2022, legateOstracismRealLife2021, wesselmannInvestigatingHowOstracizing2020}, what drives people to sever a connection (especially with close friends or partners), acting as the \emph{initiator}? On the receiving end, what are those people's feelings and reactions, as the \emph{recipient}? 

Past work investigated these questions through specific reason-driven lenses such as breakups \cite{pinterAmNeverGoing2019, pinterBeholdOnceFuture2022, pinterWorkingErasingYou2024, codutoDeleteItMove2024}, irreconcilable political views \cite{zhuPoliticalImplicationsDisconnective2024, yangPoliticsUnfriendingUser2017}, and harassment \cite{whittakerCyberbullyingSocialMedia2015, jhaverOnlineHarassmentContent2018}. Prior work has often identified the reason first and worked backwards to understand the experience. However, not every instance of \textbf{digital severance}, especially between close dyads, stems from a clear reason. 

We focus on severance as an experience, rather than a consequence, highlighting the feelings, motivations, tensions, and actions that accompany a fracture in a relationship. We define \textbf{digital severance} (or henceforth, \textbf{severance}) as an intentional, platform-mediated response to relational rupture to immediately remove digital visibility and eliminate communication channels. Although these responses lie on a spectrum, severance focuses on the active, final actions that remove both connection and availability completely --- i.e. blocking and unfriending --- creating an observable and intentional boundary \cite{wisniewskiFightingMySpace2012, karr-wisniewskiNewSocialOrder2011a}. Severance is more than an action; it is a felt response to immediately ``get rid'' of someone from one's life, contrasting with slow drifts or passive fading. To understand the experience of digital severance, we ask:

\begin{itemize}
    \item \textbf{RQ1} --- For both initiator and recipient, what are the motivations, feelings, tensions, and actions in the moments before, during, and after digital severance?
    \item \textbf{RQ2} --- How does digital severance represent broader relationship dynamics?
    \item \textbf{RQ3} --- How does our understanding of digital severance inform the ways we consider technologies around relational rupture?
\end{itemize}

Through interviews with 30 participants, we developed themes that formed a narrative around the severance experience. For almost all participants, both the relationship \emph{before} severance and the emotional processing \emph{after} involved both offline and online components. We examine \textbf{digital severance} as a distinctly digital act that arises from both offline and online experiences, enacted and made visible through social media platforms. 

While severance is meant to be a simple way to end unwanted digital connections, we find that the emotions and tensions underlying the experience can be complex and linger in the aftermath. These feelings persist, especially in the context of a previously close relationship. We find that the reasons why people take various platform-provided severance actions are a result of several hidden internal tensions, in which their actions can be viewed as both a form of personal control or relational power. We also highlight where currently available severance actions fall short in addressing each party's emotional needs, suggesting that both the recipient and initiator need support tools to deal with residual negative emotions. We contextualize our findings around platform design, social media dynamics, and well-being technology to propose open questions for designers and researchers, and provide initial ideation suggestions for future design opportunities. 

Our contributions are: (i) we define digital severance as a platform-mediated response for immediate social exclusion and how such actions can be representative of both relational power and control, (ii) we highlight the motivations, feelings, and tensions underlying severance for both initiator and recipient, and highlight how severance actions partially address and partially mismatch each party's wants, and (iii) we offer design suggestions for supporting the feelings that arise as a result of severance without erasing emotional nuance or care. 

\textbf{Content Warning}: This paper contains mention of emotional abuse.

\section{Background and Related Works}  

We examine computer-mediated relationships and social media, while understanding how communication and relationships differ on digital platforms. We then look at literature related to social exclusion and draw similarities between ostracism and digital severance. We finally consider the spectrum of disconnective digital actions to understand motivations for such behaviours. 

\subsection{Relationships and Social Media Interactions}

Online connections, e.g., on social media platforms, facilitate digital interactions, self-disclosure, and identity exploration \cite{bonettiRelationshipLonelinessSocial2010, davisFriendshipAdolescentsExperiences2012, hoodLonelinessOnlineFriendships2018}. They are an important subject in the field of computer-mediated communication (CMC), which studies human interaction over digital mediums. Traditionally, offline friendships are theorized as more intimate, due to verbal cues that support feelings of warmth and involvement \cite{chanComparisonOfflineOnline2004, chungSocialInteractionOnline2013, mittmannTikTokMyLife2022, brignalliiiImpactInternetCommunications2005}. However, Walther developed the \emph{hyperpersonal} model of CMC, which theorizes that online interactions can also be more intense, as individuals idealize the other party and fill in mental gaps \cite{walther2011theories, waltherComputerMediatedCommunicationImpersonal1996b}. When both offline and online communication methods exist in a relationship, they can work together to reinforce the connection \cite{reichFriendingIMingHanging2012}. While prior work has considered the effect of digital connections on relationship formation and sustenance, our work considers the reverse --- how relational rupture (that can happen both offline and online) leads to severance of a digital connection.   

To begin, communication and self-identity differ significantly offline and online. Qureshi-Hurst \cite{qureshi-hurstAnxietyAlienationEstrangement2022} argues that online communication decreases interaction quality because people construct themselves as artificial and idealized. People might \emph{strive} for an authentic presentation, yet may find it challenging to disclose negative experiences \cite{haimsonOnlineAuthenticityParadox2021}. Asynchronicity (time-lapse between conversations), permanence (accessibility of content), and availability (ease of accessing content) are other important differentiating factors between offline and online communication \cite{nesiTransformationAdolescentPeer2018}, with the ambiguity, asymmetry, and dynamism of social media adding complexity \cite{hoganBreakUpsLimitsEncoding2017}. Given these complexities, Amichai-Hamburger et al. \cite{amichai-hamburgerFriendshipOldConcept2013} and Golder et al. \cite{golderRhythmsSocialInteraction2007} ask what it fundamentally means to be labelled a \emph{`friend'} online. 

This is an important question, as digital platforms fundamentally change behavioural interactions, even for connected `friends'. Two patterns of relationship avoidance distinct to digital settings are ``ghosting'' and ``orbiting''. Ghosting is when one person ignores or stops communication with another without explanation. This concept has been heavily studied in modern communications \cite{kayEmpiricalAccessibleDefinition2022, collinsUnwantedUnfollowedDefining2023, pancaniRelationshipDissolutionStrategies2022a}. Research highlights ghosting's psychological impact and emotional dynamics --- ghosters can show happiness and relief but also feel guilt and regret \cite{wuWhenSilenceSpeaks2023, freedmanEmotionalExperiencesGhosting2024}, while ghostees experience hurt and face a threat to fundamental needs \cite{freedmanEmotionalExperiencesGhosting2024}. Ghosting is perpetuated by the constant avenue of communication, the element of anonymity, potential expectational mismatch, and the ease of disappearance \cite{campaioliDoubleBlueTicks2022, thomasDisappearingAgeHypervisibility2021, parkGhostingSocialRejection2024}. Orbiting is similar to ghosting, but the disengager continues to interact with the other party's online content, creating confusion arising from ambiguous signals \cite{pancaniGhostingOrbitingAnalysis2021a, pancaniRelationshipDissolutionStrategies2022a}. 

While ghosting and orbiting are responses to rupture, our work focuses on severance actions that are decisive and intentional --- a system-level disconnection that sets a definitive boundary. Contrasted against the sustained ambiguity of ghosting and orbiting, severance forces a narrative end. In Wisniewski's work \cite{wisniewskiFightingMySpace2012, karr-wisniewskiNewSocialOrder2011a}, these actions represent the most severe of the relational and interactional boundaries --- regulating who is let in and who can interact, respectively. We explore what compels users to take these definitive actions regarding communication and visibility.  

\subsection{Ostracism and Exclusion}

We tie digital severance to ostracism research, providing a basis to examine social exclusion. People may engage in ostracism behaviours --- ignoring and excluding others --- to handle poor-performing agents \cite{wirthAtimiaNewParadigm2015} or maximize group benefits \cite{tamaiOddManOut2021}. However, the negative effects of ostracism \cite{williamsOstracism2007, williamsSocialOstracism1997} are thoroughly studied \cite{williamsOstracismSocialExclusion2022}. Ostracism is a hurtful action that undermines people's sense of personal belonging, control, self-esteem, and meaningful existence \cite{williamsOstracismConsequencesCoping2011a, williamsOstracismSocialExclusion2022}. The target of ostracism feels worse as their need for relatedness is threatened \cite{lustenbergerExploringEffectsOstracism2010}; the latter ties into the core human needs as underlined by self-determination theory \cite{ryan2022self, laguardiaSelfdeterminationTheoryFundamental2008}. Carpenter \cite{Carpenter_2020} described how people feel disconnected when they are left out; Zadro and Gonsalkorale \cite{zadroHowLongDoes2006} discussed that ostracism's negative effects are exacerbated for the socially anxious, who may interpret ambiguous situations as more threatening. Amidst such negative feelings, one way of coping is self-reflection to understand one's ostracism experience \cite{williamsOstracismConsequencesCoping2011a}. 

Surprisingly, the source of ostracism may experience negative emotions too \cite{wesselmannInvestigatingHowOstracizing2020, zadroSourcesOstracismNature2014}. Some studies indicated improved feelings of control and self-esteem for the source, but Zadro and Gonsalkorale \cite{zadroSourcesOstracismNature2014} found that engaging in ostracism, especially towards a loved one, can induce feelings of disappointment and isolation. Ostracism sources may experience feelings of guilt and shame \cite{wesselmannInvestigatingHowOstracizing2020, zadroSourcesOstracismNature2014}; Legate et al. \cite{legateOstracismRealLife2021} highlighted how some may find it difficult to feel autonomy concerning actions that are understood to hurt others. Even the presence of ostracism without direct involvement can induce concern and distress through empathetic perspective-taking \cite{giesenReallyFeelYour2018a, bernsteinOstracizedWhyEffects2018}. Overall, ostracism may induce ambiguous loss --- unclear loss without resolution \cite{bossAmbiguousLossComplicated2014, bossTraumaComplicatedGrief2010, betzAmbiguousLossFamily2006a, bossAmbiguousLossTheory2007}. Even though ostracism is hurtful, prior research suggests that humans are prone to hurt those close to them \cite{millerWeAlwaysHurt1997, karremansBackCaringBeing2004}, sometimes without intent, control, or rationality \cite{pasupathiWhenHurtOthers2019}. 

Ostracism helps frame how exclusion can evoke negative psychological effects. Our findings reveal parallels between severance and ostracism regarding their effects on relationship dynamics and tensions. However, much prior ostracism literature involves ostracism taking place offline, or, if digital, through controlled lab experiments \cite{williamsCyberballProgramUse2006, williamsCyberostracismEffectsBeing2000, wirthMethodsInvestigatingSocial2016}. Real-life and digital interactions can differ in communication and expectations, and isolated experiments fail to account for the complex network of existing and past relationships. Thus, our work considers digital severance --- a form of ostracism in \emph{digital} contexts through a ``\emph{storytelling as research}'' paradigm \cite{lewisStorytellingResearchResearch2011, riessman1993narrative}. 

\subsection{Digital Severance and Digital Disposal}

Blocking, unfriending, unfollowing, etc., are active methods to sever a digital connection. These behaviours serve different purposes and motivations like combatting cyberbullying and harassment \cite{lenhartTeensKindnessCruelty2011, gashi2016unfriending, jhaverOnlineHarassmentContent2018}, setting barriers against strangers \cite{dwyerDigitalRelationshipsMySpace2007}, selectively avoiding people with opposing perspectives or annoying posting habits \cite{bayshaDividingSocialNetworks2020, shenShieldMyselfThee2019, yangPoliticsUnfriendingUser2017, zhuPoliticalImplicationsDisconnective2024, kwakFragileOnlineRelationship2011}, hiding things from specific people \cite{hanExploringBlockingBehavior2019}, or retaining personal privacy \cite{whittakerCyberbullyingSocialMedia2015}. However, much of this past work is relationship-agnostic. In the context of a closer relationship, Sibona \cite{sibonaUnfriendingFacebookContext2014} finds that friends are unfriended for making polarizing posts or for disliked offline behaviour, and Gashi and Knautz \cite{gashi2016unfriending} outline the differences in action for hiding and blocking online friends. Regarding romantic relationships, Van Ouytsel et al. \cite{vanouytselExploringRoleSocial2016} discuss how removing ex-partners from social media after a breakup forms strategies for self-regulation and demonstrates personal growth in moving on. 

However, the distributed and highly visible nature of social media is not always conducive to moving on. Pinter et al. \cite{pinterAmNeverGoing2019} examine encounters on Facebook algorithms that show potentially upsetting reminders of past relationships, and Blackburn et al. \cite{blackburnHowWillYour2023} examine how people retain and delete various digital possessions after a breakup. While research has highlighted people's various coping mechanisms to deal with sadness and confusion post-breakup \cite{mearnsCopingBreakupNegative1991, perillouxBreakingRomanticRelationships2008}, breakups nowadays often leave static traces online as artifacts that complicate emotional resolution \cite{hoganBreakUpsLimitsEncoding2017}. Pinter has heavily investigated aspects of social identity, digital management, and removal of memories in the context of breakups \cite{pinterBeholdOnceFuture2022, pinterWorkingErasingYou2024, pinterYouGotYourself2021}. This work explores how people post-breakup treat connections as possessions and create exhibitions for themselves and act as archivists or revisionists of their digital data \cite{pinterBeholdOnceFuture2022}.

Prior work in online severance actions often focuses on a reason (e.g. cyberbullying, political disagreements) first and analysis after. However, our work takes the opposite approach. Rather than focus on \emph{why} it happens, we focus on \emph{how it feels} and \emph{what it means}. We view severance as not simply a consequence, but a relational event shaped by emotions, ambiguity, power, and control. Some instances of severance are not fully rational, especially with the complexities of relationships and social platforms. Listening to stories that detail what happens before, during, and after severance provides insight into how technologies might better support these experiences. 

\section{Methods}

We conducted semi-structured interviews with participants with digital severance experiences. The interviews were largely narrative-driven, as participants recounted their relevant experiences. We focused on understanding the context of their experiences, their emotions, and so forth.

\begin{table}[!htbp]
\centering
\Description{Table displaying demographic information (age, gender) as well as severance experience and relationship with the other party, for each of the 30 participants.}
\begin{scriptsize}
\rowcolors{2}{white}{mypink2}
\begin{tabular}{c c c C{3.5cm} C{3cm} C{3cm}}

\hline
\rowcolor{mypink2}
\textbf{ID} & \textbf{Age} & \textbf{Gender} & \textbf{Severance Experience} & \textbf{Relationship} & \textbf{Platforms} \\
\hline

P1  & 21 & M  & Severing / Being Severed       & Friend                            & Instagram \\
P2  & 25 & M  & Severing Only                  & Friend                            & Facebook \\
P3  & 22 & W  & Severing / Being Severed       & Friend                            & Instagram, Twitter \\
P4  & 26 & M  & Severing / Being Severed       & Friend                            & Facebook, Instagram, Phone, Snapchat \\
P5  & 21 & W  & Severing Only                  & Romantic Partner                  & Facebook \\
P6  & 32 & M  & Severing / Being Severed       & Friend, Family, Romantic Partner  & Facebook, Instagram \\
P7  & 26 & W  & Severing / Being Severed       & Friend, Family, Romantic Partner  & Facebook, Instagram \\
P8  & 20 & NB & Severing / Being Severed       & Friend, Family, Romantic Partner  & Facebook, Instagram \\
P9  & 27 & W  & Severing / Being Severed       & Friend                            & Facebook, Instagram, Snapchat, TikTok \\
P10 & 23 & W  & Severing / Being Severed       & Friend                            & Facebook, Instagram \\
P11 & 18 & W  & Severing / Being Severed       & Friend, Family, Romantic Partner  & Instagram, Snapchat, WeChat \\
P12 & 27 & W  & Severing / Being Severed       & Friend, Romantic Partner          & Facebook, Instagram, Phone, Snapchat \\
P13 & 29 & W  & Severing / Being Severed       & Friend, Family, Romantic Partner  & Facebook, Instagram, WeChat \\
P14 & 23 & W  & Severing / Being Severed       & Friend, Romantic Partner                & Facebook, Instagram, Twitter \\
P15 & 21 & W  & Severing / Being Severed       & Friend, Romantic Partner          & Instagram \\
P16 & 36 & W  & Severing / Being Severed       & Friend                            & Facebook \\
P17 & 27 & W  & Severing / Being Severed       & Friend, Teammate                  & Facebook, Instagram \\
P18 & 20 & M  & Severing / Being Severed       & Friend                            & Instagram \\
P19 & 20 & W  & Severing / Being Severed       & Friend, Family                    & Instagram \\
P20 & 33 & W  & Severing / Being Severed       & Friend, Family, Romantic Partner, Coworker & Facebook, Instagram, WhatsApp \\
P21 & 20 & M  & Severing Only                  & Friend                            & Discord, Instagram \\
P22 & 19 & W  & Severing Only                  & Friend                            & Instagram \\
P23 & 24 & W  & Severing / Being Severed       & Romantic Partner                  & Instagram, KakaoTalk, LINE \\
P24 & 22 & M  & Severing / Being Severed       & Romantic Partner                  & Facebook, Instagram, Phone, Snapchat \\
P25 & 24 & W  & Severing / Being Severed       & Friend, Romantic Partner          & Instagram \\
P26 & 24 & W  & Severing / Being Severed       & Friend, Romantic Partner          & Facebook, Instagram, LinkedIn, Strava \\
P27 & 27 & W  & Severing Only                  & Romantic Partner                  & Facebook, Instagram, LinkedIn, Snapchat, WhatsApp \\
P28 & 26 & W  & Severing / Being Severed       & Friend, Romantic Partner          & LinkedIn, Phone, WhatsApp \\
P29 & 20 & W  & Severing / Being Severed       & Friend, Acquaintance              & Instagram \\
P30 & 24 & W  & Severing / Being Severed       & Friend, Family, Romantic Partner  & Instagram, Snapchat, TikTok \\

\hline

\end{tabular}
\end{scriptsize}
\caption{Summary of interview participants}
\label{table:demographics}
\end{table}

\subsection{Participant Recruitment} 

Participants were recruited through a posting on our institute's online paid listings board, which is globally accessible but primarily attracts a local audience. The eligibility criteria were to be age 18 or older and to have a self-reported shareable experience of digital severance with (self-reported) close relationships, whether on the severing end, the receiving end, or both. We recruited a sample of 30 participants (ages ranging from 18 to 36, mean of 24.2; gender distribution of 7 men, 22 women, and 1 non-binary participant), and collected self-reported data regarding their experience with digital severance behaviours, the relationship with the other party(ies), and the platform(s) severance took place on (see Table \ref{table:demographics}). Our sample size was deemed satisfactory regarding information richness and depth (subjectively evaluated based on Malterud et al.'s work on information power \cite{malterudSampleSizeQualitative2016a}); it also exceeds the common standard in HCI \cite{caine2016}. 

\subsection{Study Protocol}

Before the study, participants were asked to read, review, and sign a consent form relating to the study's ethics approval (obtained from the institute's ethical review board), data collection and usage, and study rights. Considering the sensitive nature of the topic, we took care in respecting the participants' responses and willingness to share; the researcher took careful consideration of the tone and sensitivities of each of the participants to tactfully guide the study. During the interview, the main focus was to understand the context of the digital severance experience(s), touching upon the participant's actions, emotions, and motivations. The interviewer asked the participant to walk them through their experiences with severance, focused on listening to the participant's story, probing on interesting points, and asking the participant to reflect on their perspectives. During this time, the interviewer asked questions to attain more detail on the experience or understand motivations, for example, \emph{``Was the act [of severance] communicated beforehand?''}, \emph{``How did it feel to be blocked or unfriended?''}, \emph{``Did you hope the other person might reach out again?''}. As each person's experience could be starkly different from the others, this portion of the interview was less structured. However, we also prepared a list of possible questions with follow-ups, relating to digital design (e.g. the ease of digital severance), online communication (e.g. expectations on social media), and artifact handling. These questions were more speculative and general to severance as a whole, such as, \emph{``How do you differentiate between blocking, unfriending, muting, ghosting someone?''}, \emph{``How do social media dynamics shape your reflections on these experiences?''}. All interviews were performed online over Zoom, and interviews lasted between 39 -- 96 minutes (average: \textasciitilde63 minutes); participants were reimbursed at \$16 per hour. 

\subsection{Methodological Limitations}

Here, we note some important methodological limitations regarding the data. In terms of participant sample, we note that the demographics of our study lean towards younger women, a possible limitation due to recruitment methods. Gender and age have been studied as salient factors that affect the experience, emotions, and expectations for close relationships \cite{macevoyWhenFriendsDisappoint2012, clarkFriendshipExpectationsFriendship1993, foxAgeGenderDimensions1985b} (and thus, potentially severance of such relationships). Furthermore, as recruitment was primarily local to a major urban city in North America, the provided perspectives and social media platforms centred around regional attitudes and popularity. Overall, this highlights potential future work in considering the effects of severance on a broader demographic.

There also existed limitations regarding our study protocol. Firstly, sometimes our questions involved participant speculation on hypotheticals, representing their imaginative thoughts rather than actual reality. Although we still found this important, we note speculation may differ from actual events. In our findings, we try to demarcate between what was actually felt and done versus what was hypothesized. Secondly, the study largely relied on the memory of past events. Some participants indicated that their memory of specific details could be hazy. However, all participants spoke quite confidently about their feelings, even if they could not remember their exact actions. Lastly, with the nature of the interview revolving around relationships and hurt caused, findings could be influenced by potential social desirability bias \cite{grimmSocialDesirabilityBias2010}, where participants may prefer to give responses to correspond positive self-presentation. 

Regarding the collected data, the study biases towards blocking experiences as blocking was the most common digital severance action amongst participants. Our findings also focus primarily on the dual perspectives of both initiator and recipient. Such findings only loosely touch upon the broader social ecosystem (e.g. mutual friends) from these perspectives. Although this larger social network was not always present in each narrative, a broader effect of severance in online social ecosystems is explorable in the future by incorporating the perspectives of mutual friends or other third parties indirectly affected by a fractured relationship. 

\subsection{Researcher Positionalities and Data Analysis}
\label{sec:positionalities}

To critically examine the data, the two primary authors performed a collaborative reflexive thematic analysis \cite{braun2021TA}. Thematic analysis allowed us to systematically identify the patterns in meaning, both inductively through the data itself and deductively through prior background research. The aspect of reflexivity was key as we aimed for an active role in data interpretation and theme generation \cite{braun2021TA}. Both researchers acknowledged their positionalities as active daily users of social media, as well as their cultural lens shaped by their upbringing in urban North America, where digital communication is nearly ubiquitous. Both researchers experienced digital severance with friends and partners, which grounded our interest in this research within personal experiences. We valued these perspectives as a guiding direction for analysis while remaining aware of biases and assumptions. Our collaborative approach allowed each of the researchers to reflect on individual positionalities and each other's biases, allowing for the proposal of alternative interpretations of data. Altogether, our analysis was informed broadly by the ontology of critical realism --- that truth is mediated by human experience, language, and culture --- and the epistemology of contextualism --- that knowledge production is contextually situated and subject to human perspective \cite{braun2021TA}. 

We familiarized ourselves with the data by listening to the audio transcripts and keeping a journal of unstructured familiarization notes. Throughout this process, we met regularly to discuss thoughts, biases, and how aspects of the data might relate to personal experiences, then engaged in an initial data coding process. Our coding approach was informed by both inductive (data-driven) and deductive orientations (theory-driven); we developed more semantic codes and more latent ones. The coding process was non-linear --- we constantly iterated and re-examined prior codes and data; the codes constantly evolved, shifted, and combined through the process. Finally, we cross-examined the codes against the data, subjectively evaluating whether they captured the diversity of meaning and satisfied a critical level of analysis. We then explored the connections between our codes to develop patterns of meaning (themes). To do so, we used visual thematic mapping techniques to generate broader categorical entities and relationships (see supplemental material). During this process, we also identified that our mapping appeared similar to a narrative structure; we took inspiration from narrative analysis studies \cite{riessman1993narrative}, and mapped our codes into a chronology of the severance experience. We collaboratively reflected upon our mapping to develop important themes that we felt were well-bounded, well-evidenced, and significant \cite{braun2021TA}; the themes paint an overarching narrative about digital severance, which we present in the following section. 

\section{Examining Platform Design}
\label{sec:platforms}

Following our interviews, we examined platform design to triangulate participant responses with observed experiences of digital severance actions. This acted as a reflective step and an audit of present trends. We observed severance actions across social media platforms identified by the participants (focusing on platforms that were mentioned more than once to highlight the most salient trends). We considered facets of design --- affordances, feedback, and ease of use --- across the entire experience of severance. This translated into general categories in an observational codebook, e.g., profile visibility and chat mediation. As some caveats, our observations explore immediate dyadic experiences; further work could consider dynamics within larger groups (e.g. group chats) or longitudinal effects (e.g. effects on the `algorithm'). After data collection, we identified design commonalities and the dimensions of severance and briefly highlight the main findings here. Platforms vary in settings and connection statuses, which complicates alignment of examined features. A simplified, shortened version is shown in Table \ref{table:platforms}, while a full observation table of features that we could immediately see during and after severance is in the supplemental material (with more detailed notes).

\begin{table}[!htbp]
\centering
\begin{scriptsize}

\setlength{\tabcolsep}{2pt} 
\renewcommand{\arraystretch}{1.25}

\rowcolors{2}{white}{mypink2}

\begin{tabular}{|
    C{1.4cm}  
    C{1.4cm}  
    |
    C{1cm}
    C{1.7cm}  
    C{2cm}   
    |
    C{1.5cm} 
    C{1.4cm}  
    C{2cm}  
    |
}

\hline

\rowcolor{mypink2}
\multicolumn{2}{|c|}{\textbf{}} &
\multicolumn{3}{c|}{\textbf{Initiator Action}} &
\multicolumn{3}{c|}{\textbf{Recipient Affordances}} \\
\hline

\rowcolor{mypink2}
\textbf{Platform} &
\textbf{Severance Action} &
\textbf{\# Clicks?} &
\textbf{Additional Confirmation?} &
\textbf{Indication of Consequence?} &
\textbf{Severance Notification?} &
\textbf{Initiator Receives Messages?} &
\textbf{View Entire Profile?} \\
\hline
Instagram & Block & 3 & Yes & Yes & No & No & No \\
Instagram & Unfollow & 3 & Yes & No & No & Yes & Yes \\
Facebook\footnotemark & Block & 3 & Yes & Yes & No & No & No \\
Facebook & Unfriend & 3 & Yes & No & No & Yes & Setting-Dependent \\
Twitter (X) & Block & 3 & Yes & Yes & No & No & Setting-Dependent \\
Twitter (X) & Unfollow & 2 & Yes & Yes & No & Yes & Setting-Dependent  \\
Snapchat & Block & 4 & Yes & No & No & No & No \\
Snapchat & Remove & 4 & Yes & No & No & Yes & Setting-Dependent \\
TikTok & Block & 3 & Yes & Yes & No & No & No \\
TikTok & Unfollow & 2 & No & No & No & Yes & Yes \\
WeChat & Block & 3 & Yes & Yes & No & No & No \\
WeChat & Delete & 3 & Yes & Yes & No & Yes & No \\
WhatsApp  & Block & 2 & Yes & Yes & No & No & No \\
\hline

\end{tabular}

\end{scriptsize}

\caption{Grouped comparison of severance affordances across platforms.}
\label{table:platforms}
\end{table}
\footnotetext{here referring to the platform, not Messenger}

\begin{itemize}

    \item \textbf{Ease and Transparency for the Initiator} --- The act of initiating severance options was straightforward on all platforms, often requiring two to four total clicks starting from the recipient's profile. The more severe and less easily reversed actions, like blocking, often also had a secondary confirmation step. In severe cases, certain platforms also informed users about the consequences of the severance action, such as the recipient not being notified or that they will be unable to message the initiator. 
    \item \textbf{Feedback of Severance for the Recipient} --- For all platforms and severance actions, recipients were not \emph{directly} notified of the action. Thus, recipients could only infer the severance action by attempting to view the initiator's profile or send a message. Even then, the feedback was sometimes ambiguous, as profiles and messaging options sometimes remained visible despite the severance action in less severe cases, complicating understanding. 
    \item \textbf{Affordances during Severance} --- We observed that severance actions primarily differentiated in severity with three primary elements: relationship status, chat mediation, and profile visibility, tying back to Wisniewski's relational and interaction boundaries \cite{karr-wisniewskiNewSocialOrder2011a}. The most restrictive, severe options (blocking on all platforms) typically impacted these three elements by removing relationship indicators, preventing messages from being received, and hiding their profile. Less restrictive actions, e.g. unfriending on Facebook or unfollowing on Instagram, altered the relationship status and could revert to default privacy settings (i.e. it was setting-dependent if the initiator's profile could be viewed), but often still maintained an avenue for continued messages. 
    \item \textbf{Reversal by the Initiator} --- Reversing severance actions that removed relationship indicators generally required re-establishment of the relationship status through a mutual, bi-directional confirmation. However, certain platforms still allowed unilateral severance reversal, such as in the case of WeChat blocking. Some platforms also implemented time delays for reversibility and re-severance; for example, Facebook transparently enforces a 48-hour delay before an unblocked contact can be re-blocked. Sometimes this delay was not transparent, for example, we observed that after removing the connection on LinkedIn, we could not immediately send a re-connection request, as it brought up an unspecified error; we only found through looking through the network response that the request gave a 400 error response of "CANT\_RESEND\_YET".  
\end{itemize}

Although this exercise explored specific platform design, we generally took an abstract view of social media platforms in our findings. We note how each platform has subtly different affordances regarding boundary management, visibility toggles, granularity of relationships, and so forth. While our findings focused on relational meaning-making and feelings as a whole across all platforms, we recognize that future work could focus more on understanding and designing for severance based on specific platform affordances.

\section{Findings}
Our findings touch upon both experiential aspects and interpretative perspectives of the data, interleaving narratives of both severance initiator and recipient. While their experiences may differ, we find that both sides share negative experiences in congruous ways, and both perspectives are crucial to understanding the intertwined meaning. 

\begin{figure*}[h]
  \centering
  \includegraphics[width=1.0\linewidth]{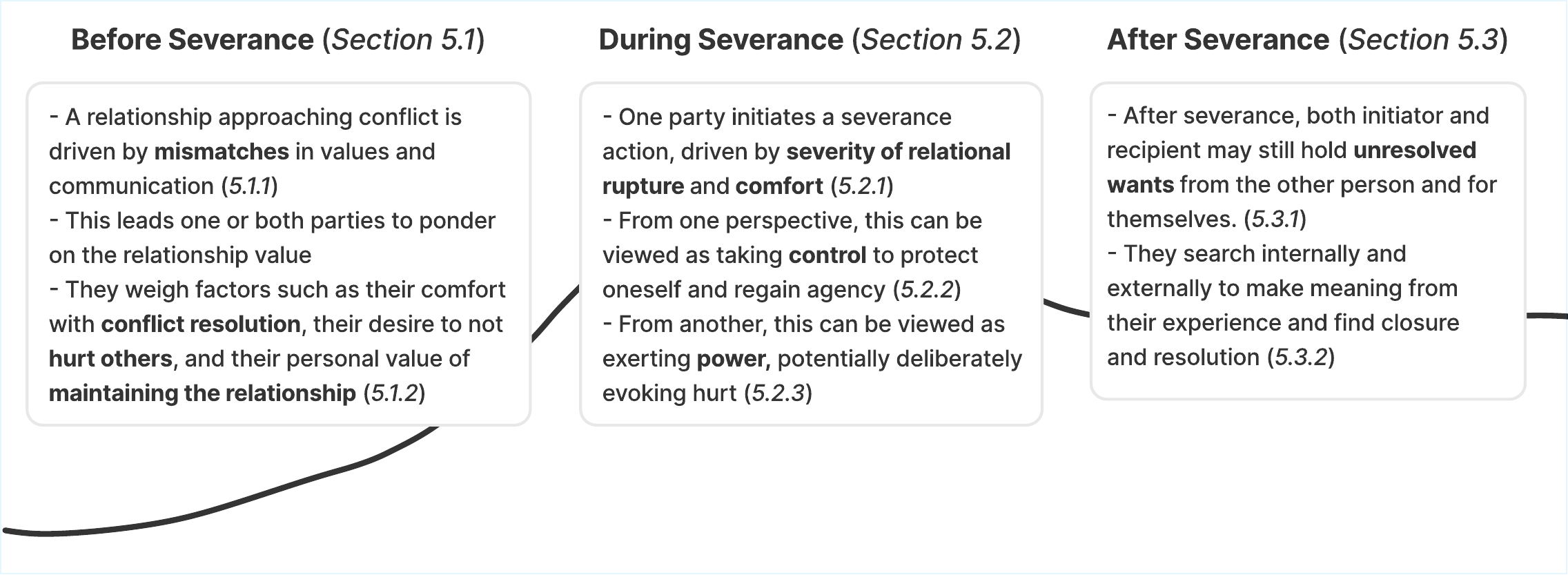}
  \caption{Summary of findings before, during, and after severance action, following the narrative shape mentioned in Section \ref{sec:positionalities}.}
  \Description{This image shows the curve-shaped story arc of severance. Starting from the left, there is a box that outlines the key findings corresponding to "Before Severance". The climax is represented by a box outlining the findings that correspond to "During Severance", and the resolution is represented by a box outlining findings corresponding to "After Severance".}
  \label{fig:summary}
\end{figure*}

Our findings address \textbf{RQ1} by exploring the experiences before, during, and after the severance action --- highlighting the changes in a relationship during the experience. We address \textbf{RQ2} by interpreting how severance relates to motifs of \emph{control} and \emph{power} in a relationship. Figure \ref{fig:summary} provides a guiding roadmap, highlighting key high-level points we expand upon in our findings. 

We focus our findings on experiences of digital severance for an \textbf{active, close} relationship, contrasting against severance that occurs when people have already drifted or when one simply wants to clean up their connections. 

\subsection{Before Severance}
We begin by highlighting the status of relationships right before severance --- understanding each side's perceptions of the relationship and the brewing tensions. 

\subsubsection{Relational Mismatches Drive Simmering Discomfort}
\label{sec:relational}
When participants discussed the problems prior to severance, we identified a common pattern of relational mismatches --- in communication and values --- that led to feelings of discomfort.

Mismatches in communication often stemmed from differing interpretations of how open communication should be. Sometimes, issues arose from clear digital violations of stated boundaries --- \emph{``I had told them I didn't want to communicate with them again''} (P12, with an ex-partner). However, these boundaries were often unclear, causing miscommunications. P23 recalls unmet communicative differences in their past situationship, \emph{``the problem with him is that he doesn't communicate and it has frustrated me''}. These communication problems escalated when left unchecked; P4 mentions that, with a friend, \emph{``I think we were gonna be bound for [severance]... I was already walking on eggshells around her''}. 

Mismatches in personality and values were another form of dichotomy in relationship dynamics that could lead to severance. Participants mentioned a shift in values could be a natural consequence of growing up and maturing, e.g. \emph{``I told him `I don't like the pessimism that you bring and I hope you understand it, but I don't really want to talk to you very much anymore'''} (P22, with a previous friend), leading to uncomfortable feelings in previously close relationships. 

Mismatches in communication and values leading to discomfort can stem from violated expectations \cite{hallFriendshipStandardsDimensions2012}, leading to an unstable relationship where one side ponders the value of maintaining it. While the relational decay could happen both online and offline, almost all participants noted how the online setting could exacerbate this perceived violation due to the permanent visibility and accessibility that is \emph{expected} online --- \emph{``if you don't perhaps receive a response instantly, you might feel like they're purposefully ignoring you''} (P7), and the lack of expression cues --- \emph{``takes a lot away from social cues and reading people's faces''} (P19). Overall, existing relational mismatches often preceded needs to sever digital ties.

\subsubsection{Accumulated Tensions Drive Potential Action}
\label{sec:accumulated}
Initial discomfort drove internal consideration of the value of a relationship. Three interconnected key tensions weighed on initiators and affected recipients in terms of actions taken to address this discomfort. We outline how these tensions inform when severance was chosen. 

\paragraph{Conflict Avoidance Versus Confronting Issues}

The first key tension was between the initiator's avoidance of discomfort in a relationship versus the initiator's comfort with confronting it. Initiators chose digital severance when they wanted to avoid conflict, and participants on both sides viewed it as a missed chance to communicate and collaboratively work through relational mismatches. Digital severance allowed one to remove someone else without communication or aggression. In some cases, this was appreciated --- P5 speculated that \emph{``It's good that you can't see if someone blocks you... It would start to feel like active aggression''}, as it can be uncomfortable to tell someone you no longer want to have a relationship with them. 

However, severance to avoid conflict was often a frustrating experience for recipients, who considered it detrimental to a continued relationship and failed to address the root of the problem. Participants described severance as impeding the resolution of problems in a repairable relationship. P17 felt that \emph{``choosing not to engage with that person anymore can be detrimental if it is a close or actively engaged relationship, where a simple conversation or phone call could solve the problem''} and P7 hypothesized that \emph{``it might take away from people developing boundary-setting skills as well as... navigating interpersonal conflict''}. The latter participant insinuated that the ease of severance on digital platforms could point towards an increasing social trend of poor conflict resolution.  
Various severance initiators also admitted that they chose severance to avoid the problem at that time. P22 states that they \emph{``didn't know how to use my words back then''} and harboured regrets over how they removed themselves from their friend's life through severance instead of working towards a resolution. Overall, digital severance was viewed as an easy escape from a now unwanted relationship, without dealing with the complexities of the underlying issues. Although the positives and negatives can be contextual, most participants felt frustrated, even regretful, about unanswered questions and unresolved conflicts with their close relationships.

\label{sec:personaldesire}
\paragraph{Personal Desire Versus Relationship Expectations}

The second key tension was between the initiator's individual wants versus expectations in the relationship. Participants understood that social exclusion is hurtful. Thus, there existed a struggle between a person's desire to sever a relationship versus their desire not to hurt the person they were close to. This was best illustrated by P15, who mentioned that they would avoid severance because \emph{``I'm just more afraid in that situation and I don't know how to communicate... you don't want to like fucking destroy someone's self-esteem''}, later adding that \emph{``I would probably have a feeling of like `I'm a bad person'''}. Severance initiators empathized with the recipient because sometimes \emph{``I feel really guilty... I know that if I were on the receiving end... I'd be really upset''} (P14). People who severed ties out of necessity wrestled with their feelings about moral propriety --- \emph{``Was [it] the right choice or not?''} (P24, with their romantic partner), \emph{``Did I make the right [or] the wrong choice?''} (P30, speaking generally about their severance experiences), needing to justify their actions to themselves or get reassurance from others.

To alleviate potential negative feelings for the recipient, some initiators attempted to communicate the end of the relationship before severance out of fairness and respect. Before severance, P21 mentioned giving notice to \emph{``two of my very close friends back in high school, right? So ignoring them wouldn't be fair''}; P30 mentions that hypothetically \emph{``if I really value that person... I would ask for that space and for that respect to be there''}. The tension between the initiator’s desire to end the relationship, while recognizing the hurt it causes, makes last instances of communication especially difficult and painful.

\label{sec:holdingon}
\paragraph{Holding On Versus Letting Go}

The final key tension was between the initiator's desire to maintain the relationship versus wanting to let go. This tension was particularly apparent in instances of breakup, which usually encoded the deepest loss. After a breakup, P5 felt \emph{``really lonely''}, recalling \emph{``I miss having someone who was really supportive all the time. So I text him and then I got annoyed by myself''}, and asking themselves \emph{``he doesn't provide emotional support that well, and so why am I still doing this?''} before blocking them online. P26 and P27 initiated digital severance of their relationships, but paradoxically also hoped for contact. P26 stated that \emph{``I was kind of hoping he was going to reach out to me, honestly''}; P27 stated that \emph{``having these multiple ways to reach gave me hope that maybe he would find some other way to reach me if he really cared''}. 

The dichotomy between holding on and letting go represented an idealized optimism versus the grounded pessimism of the situation. From the idealized perspective of wanting to stay attached: P24 reminisced about their ex --- \emph{``you stay attached to the idea of this person... You create this idealized image of a person in your eyes, and it's really easy to sort of fall victim''}. This idealization of the other party and the relationship calls back to Walther's hyperpersonal model, which theorizes that people inflate their perceptions of the other party in online communications \cite{waltherComputerMediatedCommunicationImpersonal1996b}. On the other hand, P26 highlighted how severance served as a form of letting go after a breakup. Even after the breakup, they found themselves still orbiting --- \emph{``I was checking to see if I could see an update if he was doing something... this was not a good mental space for me.''} (P26); angering and leading P26 to finality in blocking to finally let go.  

\subsection{During Severance}
As participants wrestled with various tensions, we highlight how these tensions led to specific severance actions. We also highlight how the moment of severance is related to motifs of \textbf{control} and \textbf{power}. We interpret power and control as two sides of the same coin. Control pertains to regaining digital autonomy and was largely explored from the initiator's perspective; power generally pertains to the reciprocal loss of digital autonomy and was largely explored from the recipient's perspective. Whereas control often relates to one's actions and space, power involves the influence that one has over the other person in the relationship. 

\subsubsection{Severity and Visibility Drive Selection of Severance Actions}
\label{sec:severanceactions}

Various ways exist to end a relationship online, which we found largely depended on 1) the scale of severity of relational violation and 2) the initiator's comfort with the action. The consequences of each action differ in terms of communication and visibility. Starting with strategies without severance, P22 mentioned how in such a situation, they might space out their communication and hint at their intention through tone --- \emph{``be really dry with your replies''}. When participants more explicitly began disliking the other, it escalated to ghosting --- \emph{``You don't like each other. That's why you ghost each other''} (P1, speaking generally on when they would choose specific actions). If the source of ghosting also wanted to affect visibility through notifications, then they might resort to muting as well. 

Visible severance actions (blocking, unfriending, and other active actions) were universally agreed to be more severe because they represent a definitive end in the digital relationship, as \emph{``it's very serious for someone to block and unfriend because both parties can see it right away''} (P1). Among severance actions, blocking was the most severe because it applies both \textbf{interaction} and \textbf{relational} boundaries (relating back to Section \ref{sec:platforms} and \cite{karr-wisniewskiNewSocialOrder2011a}). Digital severance is a deliberate action --- severe and retributive --- that conveys clear finality and encodes a specific digital status. Although undoing severance is possible, our observations of platform design highlight that platforms do not tend to make this possible unilaterally. Requiring the other person's participation to reconnect can create hesitation, as P5 notes that it may potentially show \emph{``you're going through an emotional roller coaster or you lack self-control''}.

Typically, severance was applied on the main communication platform, even if people were connected on others. Often enough to make the message clear, sometimes people still wanted an open line of communication due to the previous closeness, as \emph{``if he wants to reach out to me at some point, I still want him to be able to do that''} (P23, for their previous romantic partner). P1 stated that the only hypothetical exception would be to \emph{``emphasize dislike''}. Participants remarked on how simple removing someone's presence online was compared to in-person, as digital severance immediately takes effect, creating a clear loss of connection status and online interaction privileges. 

\subsubsection{Severance Represents Control for Yourself}
\label{sec:control}
Severance can provide participants with a sense of control. By limiting their own interaction and relational access, they curate what they see online. P15 expressed relief that they no longer saw their friend's posts about how \emph{``she's not doing well, she wants people to hang with her''}, P30 removed their friend online because \emph{``I don't wanna see this stuff anymore''} as seeing their previous friend's happiness evoked difficult emotions. By removing an unwanted relationship from their algorithmically generated social media feeds, participants were generally able to quell their negative feelings regarding the relationship, \emph{``just like, 'Can I forget about the whole thing and forget the whole problem?'''} (P2, after blocking their previous friend); P13 drew a metaphor of blocking actions as \emph{``clearing something out of my life''}. 

Severance also removes an account's presence on common social media platforms (e.g. Instagram or Facebook) and allows a user to regain control of the privacy of their digital space. Severance allowed participants to curate what others could see of them online, as it often removed unwanted others from encroaching on private digital space. P7 expressed their general use of severance as a tool to set \emph{``a literal boundary they cannot cross social media-wise... the power of blocking as a form of empowerment and protecting my peace''}. Extending on this point, not everyone was comfortable with letting specific friends into their digital space. P13's previous boss, with whom they had been friendly, followed them on Instagram, and P13 \emph{``felt like this had crossed the boundaries''} and blocked their boss due to wanting privacy.

Like all social boundaries, severance could be difficult to control and maintain. P7 spoke of a general \emph{``emotional barrier''} around maintaining severance. With the accessibility of severance actions making them easy to undo, this barrier was ambiguous for P27, who blocked and unblocked their situationship multiple times, being unsure about how they wanted to control their romantic relationship --- \emph{``it does take a little bit of mental strength to keep that lock there''}.

Maintaining control of oneself in a relationship (of their space and their feelings) was largely attributed to \textbf{the needs of protecting oneself and of regaining agency}. Regarding protecting oneself, severance was typically related to regulations of one's feelings. For example, P30 with an ex --- \emph{``I need to unfriend you and block you and I need space from you for a while until I'm ready to do so''}. Regarding regaining agency, severance was an easy way to reclaim control in any situation where they felt unstable and helpless. For instance, P7 drew from their psychology background, speculating that severance, especially blocking, represents a form of empowerment as \emph{``you can take control of the situation in your hands''}, thinking back to a past argument with a friend, and P26 mentioned that they unfriended and unfollowed on all socials as they \emph{``needed to be in control of something at that time... And I was just like, I need him out''} after a confusing breakup. 

Often for the initiator, increased control by one side was conversely mirrored by decreased control for the recipient. After being severed online, recipients generally felt like they were at the mercy of the other side, and their control over the relationship was lost, as P23 stated that \emph{``I cannot control his feelings''} with a romantic partner. For recipients, what they could control was the outcome and their actions. For instance, despite this disempowering feeling of being blocked by their ex, P15 accepted that \emph{``they blocked you for a reason. Doesn't matter if you know that reason or not, [you] still have to respect it''}, highlighting their way of moving on.

\subsubsection{Severance Represents Power Over the Other}
\label{sec:power}
While the motif of control examined one's relationship with themselves and their feelings, the motif of power examines one's relationship with the other. \textbf{Using severance as a deliberate tool to cause hurt or evoke negative feelings in others} consolidates power in communication. Deliberate misuse of digital severance could lead to ill-treatment teetering on emotional abuse, largely acting as a retributive action. P27 felt that their ex-situationship often disrespected them; P27, in turn, used severance as a method of retaliation to hurt them --- \emph{``At first, [blocking] was probably to punish him... I just wanted him to message and find out that I blocked him, and I just wanted him to feel some kind of sadness''}. Although most participants were unaware of the potential harm done by severance or were aware but empathetic, P27's actions were deliberate in creating harm, tied to maintaining power in the relationship. For P25 --- \emph{``I kicked her off my TikTok because if I posted a TikTok and then she saw it and then she unfollowed me or she blocked me after --- I could beat her to that... I could have had the first win by doing it first''}, implying that being the first to sever the tie was a form of ``winning'' because they could hurt their previous best friend first, further solidifying severance as a retributive action. 

From a recipient's perspective, P28 mentioned how severance can be emotionally abusive when they lost their ability to contact their previously close housemate when they were in an accident, lamenting that \emph{``people don't understand that we are not a number on their phone. We are not a profile on their social media, we are real people and it is emotional abuse''}. The housemate who cared about P28 blocked P28 --- \emph{``That lady I lived with, she would block me and unblock me only if she [had] to send me something. And then, she would block me again''}. Such actions allowed communication only on the severance initiator's terms, creating an imbalance of power within the relationship. P28 speculated that this pattern of abuse began after the housemate had felt slighted by something P28 had said, reiterating severance as a tool for revenge. 

This motivation for severance is retributive and implies that addressing perceived hurt is reciprocating hurt. P25 speculated on how \emph{``for some people if they're in a really bad situation where it's manipulative... and they're experiencing bad emotional turmoil, that can really be used to manipulate someone's feelings''}. Resolution through healthy communication can be limited as \emph{``their partner will have to resort to medieval [note: in-person] ways of reaching out''}, as further speculated by P25. Furthermore, not all relationships have these ``medieval ways'' of contact, especially long-distance relationships. Some participants hypothesized that closing avenues of communication potentially leads to drastic measures to track the other side down, exacerbating the damage to the relationship. While these extreme cases were rare, they highlight how digital severance can potentially create a power imbalance in a relationship.

\subsection{After Severance}
Finally, we consider the aftermath of severance. Our previous sections have highlighted the primarily negative emotions that can arise from severance; here, we explore the tensions, processing, and actions that happen after. 

\subsubsection{Feelings of Ambiguous Loss Drive Wants}
\label{sec:wants}
Negative emotions for both initiator and recipient often resulted in them wanting to alleviate these emotions. We highlight three categories of wants that people had --- wants from the other party (\emph{wants from them}), wants from external parties (\emph{wants from others}), and wants from yourself (\emph{wants from self}).

\emph{Wants from them} manifest as an unmet resolution upon breaking up. For example, P26 and P27 both wanted an apology after a breakup, yet \emph{``I just had to come to terms with the fact that I wasn't going to get what I wanted''} (P26) and \emph{``because I never got that, I feel there's no closure''} (P27). Severance recipients were sometimes curious why they were severed without a given explanation or opportunity to respond. This could create a depressive loop --- P4 hypothesized that had they not gotten closure in a final conversation with their friend, they might have \emph{``spiral[ed] out and just start[ed] second-guessing''}. One way of seeking closure was to accept the severance and do the same back --- \emph{``[unfollowing her back] prevented me from sitting and creeping people's profiles''} (P17, with a previous friend). Both initiators and recipients generally understood that their `wants from them' would likely remain unrealized, given the severity and visibility of severance. 

\emph{Wants from self} manifest as the desire to feel better in one's personal decisions and actions (tying back to \ref{sec:personaldesire}). Both initiators and recipients justified their actions internally to comfort themselves, even if it meant exaggerating details. Dealing with the feelings after unfriending a close friend, P22 \emph{``help[ed] myself by saying `she's doing well without me', so it was like that we were bound to drift either way''}; after blocking a situationship, P27 tried \emph{``to convince myself that I don’t know what he's really thinking... if I keep that in mind, it makes me feel better''}. Beyond reflection on the experience, the other extreme of alleviating negative emotions was through repression, as \emph{``by forgetting it, I think it's the best way for me to be mentally sane''} (P1, after mutual severance with a friend). `Wants from self' therefore represented the desire to personally feel better, while knowing that severance can hurt the other party. 

Difficulty in obtaining these wants resulted in sadness, grief, and disempowerment, making resolution and closure difficult to achieve. The last category of \emph{wants from others} helped alleviate these feelings. By talking to others --- such as friends, family, or counsellors --- people were reassured and validated. For instance, P7 talked to other friends who \emph{``understood my point of view''} after a messy disintegration of a friend group through severance by multiple parties. Speaking with others helped support feelings of resolution by aiding with reflection, resulting in lessons learnt for personal growth. For instance, P13 \emph{``went to therapy''} and said \emph{``I read books about relationships. I tried to figure out things that had happened within our relationship''} after she had been severed by a previous partner. P25 had discussions with mutual friends that changed \emph{``how I see the situation''}; allowing them to learn how their actions was hurtful to their friend leading to mutual unfollowing. Others serve as proxies to fill the void left by unachievable relationship wants and challenging personal desires. 

\subsubsection{Time, Processing, and Social Support Drive Change}
\label{sec:change}
The actions taken to move on and achieve closure, for both initiators and recipients, were largely external to the digital setting and relationship. Participants discussed actions such as connecting with others, busying themselves with work/school/hobbies, or attending counselling. These actions allowed people to regain meaning by reconnecting with other important parts of their lives. Unhealthy coping mechanisms gave way to reflection and self-growth, as P6 mentions --- \emph{``I'm trying to work on myself right now and try to find peace and then it's letting go of things''} after their previous experiences with severing relationships. Over time, the initial visceral emotions can start to fade --- time was the greatest healer as people adapted to life without the other party. 

Both initiators and recipients were able to eventually treat their painful experiences as learning ones after reflection, gaining \emph{``a lot of maturity, a lot of more patience''} (P8), and \emph{`the space and time to understand my own emotions and understand what I was feeling''} (P14). A few participants, such as P23, \emph{``went back to read messages and it helped me reflect on myself too''}. P5 states how \emph{``it's important for me to remember periods of my life... where I changed as a person''} even after blocking their previous partner. As people made sense of their experience and reflected on their insights, immediate negative emotions faded and were replaced by lessons for the future. 

As one's relationship with oneself changed, so did their feelings towards the other party. As instinctive feelings of confusion, dislike, etc., faded out, many participants (both initiators and recipients) hypothesized they would maintain cordiality if they were to meet again --- \emph{``just be cordial, and not friends per se, but acquaintances, strangers''} (P7, with her prior friend group). While many discussed severance actions that still persisted up to the time of the interview, some initiators had reversed the severance activity after some time, as \emph{``enough time has passed. This is silly. I've made peace with all of this''} (P6, with almost all of their previous severed relationships), and a few participants even expressed openness for forgiveness, if not necessarily repair --- \emph{``let bygones be bygones... if they want to talk to me, they can''} (P8, after mutually blocking a previous friend), \emph{``there's always an opportunity for rekindling of relationships''} (P12, speaking generally about relationship reconciliation after severance). 

\section{Discussion}

We acknowledge that digital severance is inherently multifaceted and context-dependent. In many cases, severance is justified and clear-cut, such as instances involving harassment or other harmful actions \cite{jhaverOnlineHarassmentContent2018, whittakerCyberbullyingSocialMedia2015}. However, compared to prior work, our findings suggest a significant gray area where blame and fault are more ambiguous, and there exists a strong emotional fallout that severance does not address for both parties. Social media platforms presently default to punishment as a simple form of safety and protection, but rarely consider what happens after. Although this may be appropriate in most cases, our findings highlight how human relationships can be complex, fluid, and ever-changing. Perhaps fortunately, digital severance encodes a discrete digital action that is easily understood and distinguished. As such, it is much easier to consider in the context of design. Here, we explore the nuances of digital severance around three key tensions, forming design guidelines to address \textbf{RQ3}. 

\subsection{Social Media Dynamics and Dyadic Online Relationships}

Whereas both parties engage in and hold feelings in a relationship, both our interview findings and observations highlight how severance is initiated and maintained by a single party. We discuss the implications of this contrast on social media communication and online relationships, highlighting the tension \textit{between the bi-directional nature of an online relationship and the uni-directional nature of severance actions}. Section \ref{sec:relational} agrees with prior research on how people have expectations in their relationships based on understanding, self-disclosure, similarity, and more \cite{cillessenPredictorsDyadicFriendship2005, hallFriendshipStandardsDimensions2012}. Relationships are \emph{co-constructed} through reciprocity of these characteristics \cite{oswaldFriendshipMaintenanceAnalysis2004}. On the other hand, violation of expectations can generate negative attitudes \cite{burgoonInterpersonalExpectationsExpectancy1993}, which can cause relationships to decay, eventually leading to severance. 

The first open question established from this tension focuses on proactively preventing relationships from entering this state of mismatched expectations and accumulated tension: \textbf{1. How can healthy communication boundaries be established and communicated in online relationships while respecting each individual's privacy, autonomy, and desires?} Whereas relationships could degrade both offline and online, participants noted in Section \ref{sec:relational} that the affordances of digital interaction exacerbate mismatches in communication and result in discomfort. This follows logically from prior CMC theory that contrasts online and in-person communication: a lack of nonverbal cues online creates ambiguity in emotions and intentions \cite{walther2011theories}, and Walther's hyperpersonal model of digital communication suggests that people may fill this gap with idealized assumptions \cite{waltherComputerMediatedCommunicationImpersonal1996b} leading to possible expectation violations \cite{burgoonInterpersonalExpectationsExpectancy1993}. These mismatches can lead to the loss of interaction \cite{wuWhenSilenceSpeaks2023}, and more severe offences can lead to severance. Severance may occur more readily in CMC due to the deindividualization of the initiator from the other party \cite{walther2011theories}. 

Overall, reciprocity in communication is important for a strong relationship, and contrasting attitudes between parties might indicate a misinterpretation of expectations rather than intentional infringement. With increased access to the Internet, users are increasingly expected to be online and available \cite{thomasDisappearingAgeHypervisibility2021}. Yet that is not true, as simply because a person is `online' may not mean that the interactional boundary of chatting exists --- but then what does being `online' really mean? The often binary status of people's availabilities on social media does little to set expectations that could prevent relationship issues from arising. Furthermore, the binary classification of `friend/follower' versus `not a friend/follower' fails to reflect the complexities of human relationships. Rather than frame this ambiguity as a personal emotional challenge, we look at how design can reduce ambiguity. 

Severance touches upon both interactional and relational boundaries set forth by Wisniewski \cite{karr-wisniewskiNewSocialOrder2011a}. We propose specific design considerations for social media platforms to better represent the complexity of these boundaries to prevent missteps. To start, providing a more diverse set of availability options (e.g. do not disturb \cite{jarupreechachanNotDisturbImplication2023}) better communicates one's immediate interactional boundary \cite{karr-wisniewskiNewSocialOrder2011a}. Users themselves (or with computer-mediated assistance) can set more granular availability indicators beyond binary states, such as ``available'', ``busy'', or ``scrolling'' to better reflect their interactional boundaries and reduce mismatches in communicative expectations. A more diverse set of relationship status indicators (e.g. `close friends' on Instagram \cite{sihombingPhenomenologyUsingInstagram2022}) better represents the diverse relational boundaries that one has with all of their online connections. As interactional boundaries and relational boundaries are more precisely represented, this can lead to more granular interaction possibilities as well. For example, we envision that the platform itself could inhibit (e.g. provide warnings before messages are sent) if they overstep the visible boundary. With more specific boundaries, we hope that interactions can occur with stronger mutual consent and more understanding of when boundaries are overstepped, possibly preventing relationships from entering conflict that could result in severance.  

The second question is an open question interrogating the severance action itself: \textbf{2. To what extent is the monadic centralization of power in online dyadic relationships permissible?} Tying back to the problematic, retributive view of severance as loss of power for the recipient (Section \ref{sec:power}), digital severance can immediately result in one party losing agency in a relationship, as visibility of the other and the means of communication are suddenly removed. We observed several cases in which one side of a long-distance relationship suddenly lost all contact with the other, leaving them disempowered and frustrated at being unable to respond and being uncertain about whether they will ever reconnect. While the flip side of control for the recipient (Section \ref{sec:control}) suggests that severance actions \emph{should} accompany limitations in interaction and communication, is it necessarily always fair for severance recipients to lose the opportunity to even try to communicate? There is a difference between immediately putting up a wall versus shouting at a void; the latter at least provides the person with a voice and a chance to get their side out. When digital severance can represent power, excessive misuse can verge on emotional abuse \cite{navarroPsychologicalCorrelatesGhosting2020, biolcatiCyberDatingAbuse2021}. Thus, what are more equitable ways to distribute agency over dyadic communication? 

We consider two possible ideas here --- one grounded in moments during severance and one in anticipation of it. For the former idea, we consider how power imbalance warnings could be detected by the system and enforced through limits. This is already in place on certain platforms. For example, a system that delays re-blocking (e.g. in Facebook's case, enforcing a 48-hour limit) prevents people from using it with negative intention, disallowing complete control over communication that participants mentioned could be used as punishment. For the latter idea, we draw upon anticipatory design principles \cite{cerejoAnticipationToolDesigning2024, zamenopoulosAnticipatoryViewDesign2007}. One proposal is to form a contractual pact about the shape of severance long before severance occurs (e.g. a pact stating that each party can leave a final message or apology before communication is cut). This would then be mediated by the platform if severance occurs, preventing the sudden, unexpected disempowerment of recipient autonomy that Section \ref{sec:wants} highlighted could potentially lead to spiralling and rumination drawn from an unsatisfactory resolution without any way to explain or reason. These represent initial ideas to provide a slightly more equitable distribution of agency at the end of a relationship. 

\subsection{Narrative Meaning-Making and Supporting Ambiguous Loss}

Extending on the motifs of control and power, digital severance can represent changes in the balance of autonomy and agency within a relationship. This shift in balance can cause both the initiator and recipient of severance to experience a wide mix of emotions depending on the context, from relief and calmness to uncertainty and grief. These feelings manifest through their unmet desires (Section \ref{sec:wants}) and resolution-seeking actions (Section \ref{sec:change}). For both parties, severance can be a disempowering experience that drives feelings akin to ambiguous loss \cite{betzAmbiguousLossFamily2006a, bossAmbiguousLossTheory2007} (the loss of a previously close relationship potentially without achieving closure) and ostracism \cite{wesselmannInvestigatingHowOstracizing2020, zadroSourcesOstracismNature2014} (engaging in and being the target of social exclusion). We explore the tension \textit{between one's feeling of agency when connected and the disempowerment when this connection is severed}. 

Ambiguous loss comes without closure and complicates the process of meaning-making \cite{betzAmbiguousLossFamily2006a, bossAmbiguousLossTheory2007, lefebvreGhostedNavigatingStrategies2020}. Closure was a major topic brought up by participants. In Section \ref{sec:wants}, people discussed \emph{wanting} to reach a resolution and adjust to their new life without the past relationship. Yet, initiators and recipients discussed their potential \emph{feelings} of non-resolution after the severance experience, balancing between holding on and letting go. Ambiguous loss created difficulty in drawing out meaning because the lack of closure confounds the narrative conclusion and the overarching thematic takeaways from the experience. This prevents people from reaching a satisfying conclusion to move on \cite{bossMythClosure2012} and creates feelings of general disempowerment \cite{bossAmbiguousLossComplicated2014}.

While social media platforms provide \emph{simple mechanisms} for relational connection and severance, they largely ignore the \emph{complex emotions} that come intertwined with relationships. Platforms have little incentive to support closure or emotional reflection, which are invisible from engagement metrics. Instead, these are largely left for the users to figure out. Yet given the negative emotions of digital severance that can potentially lead to rumination and negative spirals without support \cite{kirkegaardthomsenAssociationRuminationNegative2006, curciNegativeEmotionalExperiences2013}, we ask: \textbf{3. How can we design systems that support closure or resolution after severance, an emotionally complex and disheartening experience for both initiator and recipient?} Although we cannot force reconciliation, we recommend \emph{reflection} as a way of providing support, shifting inward exploration into outward change \cite{atkins1993reflection} and improving health, well-being, and growth \cite{bryant2005using, lyubomirsky2005pursuing, slovak2017ref-practicum}. 

Researchers have shown how sense-making efforts can help people come to terms with painful experiences and help them move on \cite{wilsonfadijiExploringMeaningMakingUniversity2022, todorovaWhatThoughtWas2021}. One common method to draw out meaning was through the development of personal narratives, as stories help with the interpretation of life events \cite{morganWhyMeaningmakingMatters2020, castiglioniFosteringReconstructionMeaning2020}. For example, journaling has often been studied as a form of systematic introspection supports meaning-making from difficult life events \cite{ullrichJournalingStressfulEvents2002c, richelleStayPositiveEffects2024a}. Through writing down their thoughts and feelings, journalers can process, express, and release their difficult emotions \cite{walkerWritingReflection1985, cowanFacilitatingReflectiveJournalling2013a}. We imagine that social media platforms (either on the platform itself, or guided to an external app) can guide both initiators and recipients to opt-in writing practices after severance. By engaging in sense-making through journaling, users can understand their feelings and experiences to help attain a sense of closure \cite{walter2008journaling}. From a digital perspective, such journals could also incorporate context-specific prompts \cite{zhouJournalAIdeEmpoweringOlder2025} to support users in walking through each step of the experience and how they felt before, during, and after severance. 

Prior work has also explored other technology-mediated ways to support closure and letting go, e.g. through rituals and tangible artifacts \cite{corina2016grief, ravn2024materializing}. Although these prior works support the experiences of the bereaved, we can appropriate these ideas to support people who endure ambiguous loss as well. These systems could help support people coming to terms with their wants (referring back to \ref{sec:wants}). For instance, directly after severance, we found that participants often wanted to expunge and forget to get rid of the other person. Digital disposal of persistent artifacts can facilitate closure and letting go; thus, the implementation and design of disposal processes could be important for future study to achieve closure and resolution \cite{gulottaDigitalArtifactsLegacy2013, pinterBeholdOnceFuture2022, pinterWorkingErasingYou2024}. For example, a digital tool to support symbolic discarding of relational artifacts \cite{wagenerMoodShaper2024}, e.g. similar to Gulotta et al.'s probes \cite{gulottaDigitalArtifactsLegacy2013}, may represent the fading lifespan of a relationship and support users in letting go of their wants after severance. 

After participants had started to engage in resolution and growth, some participants were emotionally prepared to revisit shared memories and artifacts. Here, we can draw upon various digital practices such as memory curation \cite{whitakerMakingArrangementsCuration2022} to support memorialization, metaphorical interaction \cite{wagenerMoodShaper2024} to support reframing, and alternate perspective-taking \cite{southworthBridgingCriticalThinking2022, muradovaSeeingOtherSide2021} to understand both sides of the relationship, depending on the user's diverse reflective needs. 

\subsection{Building Technology for Possible Reconciliation}

We return to the view of digital severance as potential punishment (Section \ref{sec:power}). This current section discusses both immediate and future outcomes of digital severance, focusing on the dichotomy between retribution and forgiveness after perceived transgressions. We explore the tension in technological design \textit{between initial penalization of harmful actions and future opportunities to atone}.

Human experiences operate on different time scales and constantly evolve in thought, emotion, and action \cite{wieseDynamicalSocialPsychology2010, peterson2006people}. In Section \ref{sec:change}, participants reflected upon severance actions far in the future, and often felt like their anger, sadness, and uncertainty had faded over time. People naturally let go of the past \cite{crawfordTimeHealsAll2023, walkerFadingAffectBias2009}, and some participants came to respect the positive aspects of the relationship. Given human complexity, we are prone to hurting those we are close with \cite{karremansBackCaringBeing2004}, even when unintended \cite{pasupathiWhenHurtOthers2019}. Perpetration and victimization of hurt are interleaved, and people can tell differing narratives about hurtful experiences from both sides: hurt can sometimes be purposeful, intentional, incomprehensible, or uncontrollable \cite{pasupathiWhenHurtOthers2019}. Participant narratives revealed that digital severance experiences in close relationships can both derive from and create feelings of hurt \cite{vangelistiReactionsMessagesThat1998}. 

Instead of addressing these feelings, online severance encodes a near-permanent disconnection of that relationship, with both sides uncomfortable reaching out afterwards due to the clear and visible division. Although the researchers have personally seen reconciliation occur, it is somewhat telling that essentially every severed relationship from the study remained fractured. While unblocking and re-adding previously severed relationships is possible, this usually requires the reformation of a bi-directional contract, e.g., re-friending. Not unlike severance, the meaning of these reconciliation actions can be ambiguous and speculative --- does it signal forgiveness, moving on, openness for reconciliation? Although some participants spoke about potential reconciliation and cordial interaction after severance reversal in Section \ref{sec:change}, none actually discussed taking action to communicate this. Without such communication, how can the status of acceptable communication be known? Even before that, which party might even be inclined to reach out first? 

A complete disconnection is possibly the worst ending for a previously close relationship. Prior research has argued that relationship resolution can help facilitate empathy and resolution \cite{vasalouApplicationForgivenessSocial2009}; thus, given the potential for human change and reflection in the aftermath of hurtful actions: \textbf{4. How can we design to allow for potential reconciliation, forgiveness, or atonement?} 

Social media platforms rarely have incentives to consider what happens to both parties after severance --- they are a `system whose design is dominated by tools of punishment' \cite{vasalouApplicationForgivenessSocial2009}, especially given the ease of severance actions. Yet, as people can change, reflect on their hurtful actions, and recognize their mistakes, how can technology potentially reconcile the severed relationship? There are guidelines from prior research that should be respected, such that forgiveness is not mandatory, is not unconditional, and may not necessarily repair trust \cite{vasalouApplicationForgivenessSocial2009}. However, an increased emphasis on restorative justice could relieve people's guilt and shame while improving their future outlook \cite{vasalouPraiseForgivenessWays2008, kannabiras2021sorry}, as a few participants from our study held onto guilt far after the severance experience. Even if forgiveness is not entirely achieved, the possibility for \emph{personal atonement} could still provide a way for people to help themselves. 

We propose a number of suggestions that could support reconciliation and atonement. For instance, apologies are often the first step in resolving a transgression in close relationships \cite{schumannDoesLoveMean2012, lewisApologiesCloseRelationships2015}. From our findings, several participants wanted an apology to help them with closure, while other participants expressed regret over their behaviours. Apologies in the real world often require a bi-directional interaction \cite{lewisApologiesCloseRelationships2015} and their outcomes are contingent on a number of factors. Given that severance is a monadic action that can be divisive across feelings of reconciliation, how can digital platforms support one-directional apologies while respecting the other party's interactional boundaries and distance \cite{karr-wisniewskiNewSocialOrder2011a}? 

Inspired by prior ideas incorporating restorative justice \cite{kannabiras2021sorry, xiaoAddressingInterpersonalHarm2023a, kieferApologyRestitutionRoleplay2020}, we propose unidirectional forms of apology-writing and atonement on digital platforms. For instance, after being severed, we propose that either party can write an apology to the other, which is never seen or notified to the other party \emph{unless} the other party takes specific steps to check for it. For example, the other party would have to navigate specifically to the message, note an icon of potential apology, and then decide whether or not they want to see it --- this design would maintain control of relational and interactional boundaries \cite{karr-wisniewskiNewSocialOrder2011a}. Even if the apology is never read, we hope that the act of expressive writing in the apology would support meaning-making and emotional release \cite{romeroWritingWrongsPromoting2008}. 

Furthermore, extending AI-assisted designs for emotional support and reconciliation \cite{liuArtificialIntelligencePerceived2024, battistiSecondPersonAuthenticityMediating2025}, we imagine that AI could serve as a neutral interlocutor (taking the position of, e.g. a mutual friend), that listens and supports reflection on both sides privately, and eventually can slowly come to share reflections, apologies, or atonement with mutual consent.

\section{Conclusion}
We interviewed 30 participants who had experience with digital severance in previously close relationships to understand how people perceive, describe, and ascribe meaning to their experiences. Through reflexive thematic analysis, we highlighted themes revealing the multifaceted nature of digital severance, finding that the digital nature of online interactions often clashes with the analog nature of real-life relationships. While the simple, binary action of severance can be beneficial, it can also perpetuate conflict avoidance, prevent reconciliation, and induce feelings of ambiguous loss. We interleaved digital severance with motifs of control and power and emphasized how its current digital design may fail to account for many gray areas regarding the complexities of human relationships. Finally, we raise open questions to spark future discussion on how to design ways of supporting both the maintenance and dissolution of relationships in online spaces.



\bibliographystyle{ACM-Reference-Format}
\bibliography{main}

@article{qureshi-hurstAnxietyAlienationEstrangement2022,
  title = {Anxiety, Alienation, and Estrangement in the Context of Social Media},
  author = {{Qureshi-Hurst}, Emily},
  year = {2022},
  month = sep,
  journal = {Religious Studies},
  volume = {58},
  number = {3},
  pages = {522--533},
  publisher = {{Cambridge University Press}},
  issn = {0034-4125, 1469-901X},
  doi = {10.1017/S0034412521000093},
  urldate = {2024-02-05},
  langid = {english}
}

@incollection{grimmSocialDesirabilityBias2010,
  title = {Social {{Desirability Bias}}},
  booktitle = {Wiley {{International Encyclopedia}} of {{Marketing}}},
  author = {Grimm, Pamela},
  year = {2010},
  publisher = {John Wiley \& Sons, Ltd},
  doi = {10.1002/9781444316568.wiem02057},
  urldate = {2025-11-13},
  copyright = {Copyright {\copyright} 2011 John Wiley \& Sons, Ltd. All rights reserved.},
  isbn = {978-1-4443-1656-8},
  langid = {english}
}

@article{kieferApologyRestitutionRoleplay2020,
  title = {Apology and {{Restitution}} in a {{Role-play Restorative Justice Experiment}}: {{Multiple Perspectives}}, {{Multiple Measures}}},
  shorttitle = {Apology and {{Restitution}} in a {{Role-play Restorative Justice Experiment}}},
  author = {Kiefer, Rebecca P. and WorthingtonJr., Everett L. and Wenzel, Michael and Woodyatt, Lydia and Berry, Jack W.},
  year = {2020},
  month = jun,
  journal = {Journal of Psychology and Theology},
  volume = {48},
  number = {2},
  pages = {105--117},
  publisher = {SAGE Publications Ltd},
  issn = {0091-6471},
  doi = {10.1177/0091647120911114},
  urldate = {2025-11-13},
  langid = {english}
}

@incollection{millerWeAlwaysHurt1997,
  title = {We {{Always Hurt}} the {{Ones We Love}}},
  booktitle = {Aversive {{Interpersonal Behaviors}}},
  author = {Miller, Rowland S.},
  editor = {Kowalski, Robin M.},
  year = {1997},
  pages = {11--29},
  publisher = {Springer US},
  address = {Boston, MA},
  doi = {10.1007/978-1-4757-9354-3_2},
  urldate = {2025-11-13},
  isbn = {978-1-4757-9354-3},
  langid = {english}
}

@article{xiaoAddressingInterpersonalHarm2023a,
  title = {Addressing {{Interpersonal Harm}} in {{Online Gaming Communities}}: {{The Opportunities}} and {{Challenges}} for a {{Restorative Justice Approach}}},
  shorttitle = {Addressing {{Interpersonal Harm}} in {{Online Gaming Communities}}},
  author = {Xiao, Sijia and Jhaver, Shagun and Salehi, Niloufar},
  year = {2023},
  month = sep,
  journal = {ACM Trans. Comput.-Hum. Interact.},
  volume = {30},
  number = {6},
  pages = {83:1--83:36},
  issn = {1073-0516},
  doi = {10.1145/3603625},
  urldate = {2025-11-13}
}

@misc{hoganBreakUpsLimitsEncoding2017,
  type = {{{SSRN Scholarly Paper}}},
  title = {Break-{{Ups}} and the {{Limits}} of {{Encoding Love}}},
  author = {Hogan, Bernie},
  year = {2017},
  month = mar,
  number = {3656516},
  address = {{Rochester, NY}},
  doi = {10.2139/ssrn.3656516},
  urldate = {2023-12-08},
  langid = {english}
}

@article{williamsOstracismSocialExclusion2022,
  title = {Ostracism and Social Exclusion: {{Implications}} for Separation, Social Isolation, and Loss},
  shorttitle = {Ostracism and Social Exclusion},
  author = {Williams, Kipling D. and Nida, Steve A.},
  year = {2022},
  month = oct,
  journal = {Current Opinion in Psychology},
  volume = {47},
  pages = {101353},
  issn = {2352-250X},
  doi = {10.1016/j.copsyc.2022.101353},
  urldate = {2024-06-01}
}

@article{whittakerCyberbullyingSocialMedia2015,
  title = {Cyberbullying {{Via Social Media}}},
  author = {Whittaker, Elizabeth and Kowalski, Robin M.},
  year = {2015},
  month = jan,
  journal = {Journal of School Violence},
  volume = {14},
  number = {1},
  pages = {11--29},
  publisher = {Routledge},
  issn = {1538-8220},
  doi = {10.1080/15388220.2014.949377},
  urldate = {2023-12-08}
}

@incollection{wirthMethodsInvestigatingSocial2016,
  title = {Methods for {{Investigating Social Exclusion}}},
  booktitle = {Social {{Exclusion}}: {{Psychological Approaches}} to {{Understanding}} and {{Reducing Its Impact}}},
  author = {Wirth, James H.},
  editor = {Riva, Paolo and Eck, Jennifer},
  year = {2016},
  pages = {25--47},
  publisher = {Springer International Publishing},
  address = {Cham},
  doi = {10.1007/978-3-319-33033-4_2},
  urldate = {2024-06-01},
  isbn = {978-3-319-33033-4},
  langid = {english}
}

@phdthesis{Carpenter_2020, series={Electronic Theses and Dissertations (ETDs) 2008+}, title={Impactful interactions: how the experience of social ostracization influences our moral judgments of others}, url={https://open.library.ubc.ca/collections/ubctheses/24/items/1.0394286}, DOI={http://dx.doi.org/10.14288/1.0394286}, school={University of British Columbia}, author={Carpenter, Tara Louise}, year={2020}, collection={Electronic Theses and Dissertations (ETDs) 2008+}}

@article{lustenbergerExploringEffectsOstracism2010,
  title = {Exploring the {{Effects}} of {{Ostracism}} on {{Performance}} and {{Intrinsic Motivation}}},
  author = {Lustenberger, Donald E. and Jagacinski, Carolyn M.},
  year = {2010},
  month = aug,
  journal = {Human Performance},
  volume = {23},
  number = {4},
  pages = {283--304},
  publisher = {{Routledge}},
  issn = {0895-9285},
  doi = {10.1080/08959285.2010.501046},
  urldate = {2024-02-26}
}

@incollection{ryan2022self,
  title={Self-determination theory},
  author={Ryan, Richard M and Deci, Edward L},
  booktitle={Encyclopedia of quality of life and well-being research},
  pages={1--7},
  year={2022},
  publisher={Springer}
}

@article{zadroSourcesOstracismNature2014,
  title = {Sources of {{Ostracism}}: {{The Nature}} and {{Consequences}} of {{Excluding}} and {{Ignoring Others}}},
  shorttitle = {Sources of {{Ostracism}}},
  author = {Zadro, Lisa and Gonsalkorale, Karen},
  year = {2014},
  journal = {Current Directions in Psychological Science},
  volume = {23},
  number = {2},
  eprint = {44318729},
  eprinttype = {jstor},
  pages = {93--97},
  publisher = {{[Association for Psychological Science, Sage Publications, Inc.]}},
  issn = {0963-7214},
  urldate = {2024-02-18},
}

@incollection{wesselmannInvestigatingHowOstracizing2020,
  title = {Investigating How Ostracizing Others Affects One's Self-Concept},
  booktitle = {Emerging {{Perspectives}} on {{Self}} and {{Identity}}},
  author = {Wesselmann, Eric D., James H. Wirth},
  year = {2020},
  publisher = {{Routledge}},
  isbn = {978-0-429-33115-2}
}

@article{zadroHowLongDoes2006,
  title = {How Long Does It Last? {{The}} Persistence of the Effects of Ostracism in the Socially Anxious},
  shorttitle = {How Long Does It Last?},
  author = {Zadro, Lisa and Boland, Catherine and Richardson, Rick},
  year = {2006},
  month = sep,
  journal = {Journal of Experimental Social Psychology},
  volume = {42},
  number = {5},
  pages = {692--697},
  issn = {0022-1031},
  doi = {10.1016/j.jesp.2005.10.007},
  urldate = {2024-02-26}
}

@article{tamaiOddManOut2021,
  title = {Odd Man out for Everyone: {{The}} Justification of Ostracism to Maximize the Whole Group's Benefits},
  shorttitle = {Odd Man out for Everyone},
  author = {Tamai, Ryuichi and Igarashi, Tasuku},
  year = {2021},
  journal = {European Journal of Social Psychology},
  volume = {51},
  number = {2},
  pages = {213--221},
  issn = {1099-0992},
  doi = {10.1002/ejsp.2725},
  urldate = {2024-02-18},
  copyright = {{\copyright} 2020 John Wiley \& Sons, Ltd.},
  langid = {english}
}

@article{wirthAtimiaNewParadigm2015,
  title = {Atimia: {{A New Paradigm}} for {{Investigating How Individuals Feel When Ostracizing Others}}},
  shorttitle = {Atimia},
  author = {Wirth, James H. and Bernstein, Michael J. and LeRoy, Angie S.},
  year = {2015},
  month = sep,
  journal = {The Journal of Social Psychology},
  volume = {155},
  number = {5},
  pages = {497--514},
  publisher = {{Routledge}},
  issn = {0022-4545},
  doi = {10.1080/00224545.2015.1060934},
  urldate = {2024-02-18},
  pmid = {26267130},
}

@article{legateOstracismRealLife2021,
  title = {Ostracism in {{Real Life}}: {{Evidence That Ostracizing Others Has Costs}}, {{Even When It Feels Justified}}},
  shorttitle = {Ostracism in {{Real Life}}},
  author = {Legate, Nicole and Weinstein, Netta and Ryan, Richard M.},
  year = {2021},
  month = jul,
  journal = {Basic and Applied Social Psychology},
  volume = {43},
  number = {4},
  pages = {226--238},
  publisher = {{Routledge}},
  issn = {0197-3533},
  doi = {10.1080/01973533.2021.1927038},
  urldate = {2024-02-18}
}

@article{giesenReallyFeelYour2018a,
  title = {Do {{I Really Feel Your Pain}}? {{Comparing}} the {{Effects}} of {{Observed}} and {{Personal Ostracism}}},
  shorttitle = {Do {{I Really Feel Your Pain}}?},
  author = {Giesen, Anna and Echterhoff, Gerald},
  year = {2018},
  month = apr,
  journal = {Personality and Social Psychology Bulletin},
  volume = {44},
  number = {4},
  pages = {550--561},
  publisher = {{SAGE Publications Inc}},
  issn = {0146-1672},
  doi = {10.1177/0146167217744524},
  urldate = {2024-02-26},
  langid = {english}
}

@inproceedings{dwyerDigitalRelationshipsMySpace2007,
  title = {Digital {{Relationships}} in the "{{MySpace}}" {{Generation}}: {{Results From}} a {{Qualitative Study}}},
  shorttitle = {Digital {{Relationships}} in the "{{MySpace}}" {{Generation}}},
  booktitle = {2007 40th {{Annual Hawaii International Conference}} on {{System Sciences}} ({{HICSS}}'07)},
  author = {Dwyer, Catherine},
  year = {2007},
  month = jan,
  pages = {19--19},
  issn = {1530-1605},
  doi = {10.1109/HICSS.2007.176},
  urldate = {2024-02-18}
}

@article{bayshaDividingSocialNetworks2020,
  title = {Dividing Social Networks: {{Facebook}} Unfriending, Unfollowing, and Blocking in Turbulent Political Times},
  shorttitle = {Dividing Social Networks},
  author = {Baysha, Olga},
  year = {2020},
  month = may,
  journal = {Russian Journal of Communication},
  volume = {12},
  number = {2},
  pages = {104--120},
  publisher = {{Routledge}},
  issn = {1940-9419},
  doi = {10.1080/19409419.2020.1773911},
  urldate = {2023-12-08}
}

@incollection{shenShieldMyselfThee2019,
  title = {I {{Shield Myself From Thee}}: {{Selective Avoidance}} on {{Social Media During Political Protests}}},
  shorttitle = {I {{Shield Myself From Thee}}},
  booktitle = {Digital {{Politics}}: {{Mobilization}}, {{Engagement}} and {{Participation}}},
  author = {Qinfeng Zhu, Marko Skoric, Fei Shen },
  year = {2019},
  publisher = {{Routledge}},
  isbn = {978-0-429-45995-5}
}

@article{yangPoliticsUnfriendingUser2017,
  title = {The Politics of ``{{Unfriending}}'': {{User}} Filtration in Response to Political Disagreement on Social Media},
  shorttitle = {The Politics of ``{{Unfriending}}''},
  author = {Yang, JungHwan and Barnidge, Matthew and Rojas, Hernando},
  year = {2017},
  month = may,
  journal = {Computers in Human Behavior},
  volume = {70},
  pages = {22--29},
  issn = {0747-5632},
  doi = {10.1016/j.chb.2016.12.079},
  urldate = {2023-12-08}
}

@inproceedings{hanExploringBlockingBehavior2019,
  title = {Exploring the {{Blocking Behavior Between Young Adults}} and {{Parents}} on {{WeChat Moments}}},
  booktitle = {Human {{Aspects}} of {{IT}} for the {{Aged Population}}. {{Social Media}}, {{Games}} and {{Assistive Environments}}},
  author = {Han, Wenting and Zhao, Yuxiang (Chris) and Zhu, Qinghua},
  editor = {Zhou, Jia and Salvendy, Gavriel},
  year = {2019},
  series = {Lecture {{Notes}} in {{Computer Science}}},
  pages = {65--76},
  publisher = {{Springer International Publishing}},
  address = {{Cham}},
  doi = {10.1007/978-3-030-22015-0_5},
  isbn = {978-3-030-22015-0},
  langid = {english}
}

@techreport{lenhartTeensKindnessCruelty2011,
  title = {Teens, {{Kindness}} and {{Cruelty}} on {{Social Network Sites}}: {{How American Teens Navigate}} the {{New World}} of "{{Digital Citizenship}}"},
  shorttitle = {Teens, {{Kindness}} and {{Cruelty}} on {{Social Network Sites}}},
  author = {Lenhart, Amanda and Madden, Mary and Smith, Aaron and Purcell, Kristen and Zickuhr, Kathryn and Rainie, Lee},
  year = {2011},
  month = nov,
  journal = {Pew Internet \& American Life Project},
  institution = {{Pew Internet \& American Life Project}},
  urldate = {2023-12-08},
  langid = {english}
}

@article{mearnsCopingBreakupNegative1991,
  title = {Coping with a Breakup: {{Negative}} Mood Regulation Expectancies and Depression Following the End of a Romantic Relationship},
  shorttitle = {Coping with a Breakup},
  author = {Mearns, Jack},
  year = {1991},
  journal = {Journal of Personality and Social Psychology},
  volume = {60},
  number = {2},
  pages = {327--334},
  publisher = {{American Psychological Association}},
  address = {{US}},
  issn = {1939-1315},
  doi = {10.1037/0022-3514.60.2.327}
}

@article{perillouxBreakingRomanticRelationships2008,
  title = {Breaking up {{Romantic Relationships}}: {{Costs Experienced}} and {{Coping Strategies Deployed}}},
  shorttitle = {Breaking up {{Romantic Relationships}}},
  author = {Perilloux, Carin and Buss, David M.},
  year = {2008},
  month = jan,
  journal = {Evolutionary Psychology},
  volume = {6},
  number = {1},
  pages = {147470490800600119},
  publisher = {{SAGE Publications Inc}},
  issn = {1474-7049},
  doi = {10.1177/147470490800600119},
  urldate = {2024-02-26},
  langid = {english}
}

@article{lewisStorytellingResearchResearch2011,
  title = {Storytelling as {{Research}}/{{Research}} as {{Storytelling}}},
  author = {Lewis, Patrick J.},
  year = {2011},
  month = jul,
  journal = {Qualitative Inquiry},
  volume = {17},
  number = {6},
  pages = {505--510},
  publisher = {{SAGE Publications Inc}},
  issn = {1077-8004},
  doi = {10.1177/1077800411409883},
  urldate = {2023-10-11},
  langid = {english}
}

@incollection{zhuPoliticalImplicationsDisconnective2024,
  title = {Political Implications of Disconnective Practices on Social Media: Unfriending, Unfollowing, and Blocking},
  shorttitle = {Political Implications of Disconnective Practices on Social Media},
  booktitle = {Research {{Handbook}} on {{Social Media}} and {{Society}}},
  author = {Zhu, Qinfeng},
  year = {2024},
  month = jan,
  pages = {135--147},
  publisher = {Edward Elgar Publishing},
  urldate = {2024-02-05},
  chapter = {Research Handbook on Social Media and Society},
  isbn = {978-1-80037-705-9},
  langid = {english}
}

@inproceedings{sibonaUnfriendingFacebookContext2014,
  title = {Unfriending on {{Facebook}}: {{Context Collapse}} and {{Unfriending Behaviors}}},
  shorttitle = {Unfriending on {{Facebook}}},
  booktitle = {2014 47th {{Hawaii International Conference}} on {{System Sciences}}},
  author = {Sibona, Christopher},
  year = {2014},
  month = jan,
  pages = {1676--1685},
  issn = {1530-1605},
  doi = {10.1109/HICSS.2014.214},
  urldate = {2024-01-11}
}

@article{bernsteinOstracizedWhyEffects2018,
  title = {Ostracized but Why? {{Effects}} of Attributions and Empathy on Connecting with the Socially Excluded},
  shorttitle = {Ostracized but Why?},
  author = {Bernstein, Michael J. and Chen, Zhansheng and Poon, Kai-Tak and Benfield, Jacob A. and Ng, Henry K. S.},
  year = {2018},
  month = aug,
  journal = {PLOS ONE},
  volume = {13},
  number = {8},
  pages = {e0201183},
  publisher = {Public Library of Science},
  issn = {1932-6203},
  doi = {10.1371/journal.pone.0201183},
  urldate = {2024-06-01},
  langid = {english}
}

@inproceedings{codutoDeleteItMove2024,
  title = {"{{Delete}} It and {{Move On}}": {{Digital Management}} of {{Shared Sexual Content}} after a {{Breakup}}},
  shorttitle = {"{{Delete}} It and {{Move On}}"},
  booktitle = {Proceedings of the {{CHI Conference}} on {{Human Factors}} in {{Computing Systems}}},
  author = {Coduto, Kathryn D and McDonald, Allison},
  year = {2024},
  month = may,
  series = {{{CHI}} '24},
  pages = {1--16},
  publisher = {Association for Computing Machinery},
  address = {New York, NY, USA},
  doi = {10.1145/3613904.3642722},
  urldate = {2024-05-30},
  isbn = {9798400703300}
}

@article{laguardiaSelfdeterminationTheoryFundamental2008,
  title = {Self-Determination Theory as a Fundamental Theory of Close Relationships},
  author = {La Guardia, Jennifer G. and Patrick, Heather},
  year = {2008},
  journal = {Canadian Psychology / Psychologie canadienne},
  volume = {49},
  number = {3},
  pages = {201--209},
  publisher = {Educational Publishing Foundation},
  address = {US},
  issn = {1878-7304},
  doi = {10.1037/a0012760}
}

@article{pinterBeholdOnceFuture2022,
  title = {Behold the {{Once}} and {{Future Me}}: {{Online Identity After}} the {{End}} of a {{Romantic Relationship}}},
  shorttitle = {Behold the {{Once}} and {{Future Me}}},
  author = {Pinter, Anthony T. and Brubaker, Jed R.},
  year = {2022},
  month = nov,
  journal = {Proceedings of the ACM on Human-Computer Interaction},
  volume = {6},
  number = {CSCW2},
  pages = {372:1--372:35},
  doi = {10.1145/3555097},
  urldate = {2024-05-31}
}

@article{pinterWorkingErasingYou2024,
  title = {I'm {{Working}} on {{Erasing You}}, {{Just Don}}'t {{Have}} the {{Proper Tools}}: {{Supporting Online Identity Management After}} the {{End}} of {{Romantic Relationships}}},
  shorttitle = {I'm {{Working}} on {{Erasing You}}, {{Just Don}}'t {{Have}} the {{Proper Tools}}},
  author = {Pinter, Anthony T. and Brubaker, Jed R.},
  year = {2024},
  month = apr,
  journal = {Proceedings of the ACM on Human-Computer Interaction},
  volume = {8},
  number = {CSCW1},
  pages = {66:1--66:32},
  doi = {10.1145/3637343},
  urldate = {2024-05-31}
}

@article{haimsonOnlineAuthenticityParadox2021,
  title = {The {{Online Authenticity Paradox}}: {{What Being}} "{{Authentic}}" on {{Social Media Means}}, and {{Barriers}} to {{Achieving It}}},
  shorttitle = {The {{Online Authenticity Paradox}}},
  author = {Haimson, Oliver L. and Liu, Tianxiao and Zhang, Ben Zefeng and Corvite, Shanley},
  year = {2021},
  month = oct,
  journal = {Proceedings of the ACM on Human-Computer Interaction},
  publisher = {ACMPUB27New York, NY, USA},
  doi = {10.1145/3479567},
  urldate = {2024-06-03},
  langid = {english}
}

@inproceedings{pinterYouGotYourself2021,
  title = {You {{Got Yourself A Whole New Life}}, and {{All I}}'ve {{Got}} Is {{Half This Old One}}: {{Breaking Up}} and {{Moving On}} in the {{Social Media Age}}},
  shorttitle = {You {{Got Yourself A Whole New Life}}, and {{All I}}'ve {{Got}} Is {{Half This Old One}}},
  booktitle = {Companion {{Publication}} of the 2021 {{Conference}} on {{Computer Supported Cooperative Work}} and {{Social Computing}}},
  author = {Pinter, Anthony T.},
  year = {2021},
  month = oct,
  series = {{{CSCW}} '21 {{Companion}}},
  pages = {283--286},
  publisher = {Association for Computing Machinery},
  address = {New York, NY, USA},
  doi = {10.1145/3462204.3481795},
  urldate = {2024-05-30},
  isbn = {978-1-4503-8479-7}
}

@article{williamsCyberballProgramUse2006,
  title = {Cyberball: {{A}} Program for Use in Research on Interpersonal Ostracism and Acceptance},
  shorttitle = {Cyberball},
  author = {Williams, Kipling D. and Jarvis, Blair},
  year = {2006},
  month = feb,
  journal = {Behavior Research Methods},
  volume = {38},
  number = {1},
  pages = {174--180},
  issn = {1554-3528},
  doi = {10.3758/BF03192765},
  urldate = {2024-06-01},
  langid = {english}
}

@article{williamsCyberostracismEffectsBeing2000,
  title = {Cyberostracism: {{Effects}} of Being Ignored over the {{Internet}}},
  shorttitle = {Cyberostracism},
  author = {Williams, Kipling D. and Cheung, Christopher K. T. and Choi, Wilma},
  year = {2000},
  journal = {Journal of Personality and Social Psychology},
  volume = {79},
  number = {5},
  pages = {748--762},
  publisher = {American Psychological Association},
  address = {US},
  issn = {1939-1315},
  doi = {10.1037/0022-3514.79.5.748}
}

@article{williamsOstracism2007,
  title = {Ostracism},
  author = {Williams, Kipling D.},
  year = {2007},
  month = jan,
  journal = {Annual Review of Psychology},
  volume = {58},
  number = {Volume 58, 2007},
  pages = {425--452},
  publisher = {Annual Reviews},
  issn = {0066-4308, 1545-2085},
  doi = {10.1146/annurev.psych.58.110405.085641},
  urldate = {2024-06-01},
  langid = {english}
}

@article{williamsOstracismConsequencesCoping2011a,
  title = {Ostracism: {{Consequences}} and {{Coping}}},
  shorttitle = {Ostracism},
  author = {Williams, Kipling D. and Nida, Steve A.},
  year = {2011},
  month = apr,
  journal = {Current Directions in Psychological Science},
  volume = {20},
  number = {2},
  pages = {71--75},
  publisher = {SAGE Publications Inc},
  issn = {0963-7214},
  doi = {10.1177/0963721411402480},
  urldate = {2024-06-01},
  langid = {english}
}

@incollection{williamsSocialOstracism1997,
  title = {Social {{Ostracism}}},
  booktitle = {Aversive {{Interpersonal Behaviors}}},
  author = {Williams, Kipling D.},
  editor = {Kowalski, Robin M.},
  year = {1997},
  pages = {133--170},
  publisher = {Springer US},
  address = {Boston, MA},
  doi = {10.1007/978-1-4757-9354-3_7},
  urldate = {2024-06-01},
  isbn = {978-1-4757-9354-3},
  langid = {english}
}

@article{vanouytselExploringRoleSocial2016,
  title = {Exploring the Role of Social Networking Sites within Adolescent Romantic Relationships and Dating Experiences},
  author = {Van Ouytsel, Joris and Van Gool, Ellen and Walrave, Michel and Ponnet, Koen and Peeters, Emilie},
  year = {2016},
  month = feb,
  journal = {Computers in Human Behavior},
  volume = {55},
  pages = {76--86},
  issn = {0747-5632},
  doi = {10.1016/j.chb.2015.08.042},
  urldate = {2023-12-08}
}

@article{pinterAmNeverGoing2019,
  title = {"{{Am I Never Going}} to {{Be Free}} of {{All This Crap}}?": {{Upsetting Encounters}} with {{Algorithmically Curated Content About Ex-Partners}}},
  shorttitle = {"{{Am I Never Going}} to {{Be Free}} of {{All This Crap}}?},
  author = {Pinter, Anthony T. and Jiang, Jialun Aaron and Gach, Katie Z. and Sidwell, Melanie M. and Dykes, James E. and Brubaker, Jed R.},
  year = {2019},
  month = nov,
  journal = {Proceedings of the ACM on Human-Computer Interaction},
  volume = {3},
  number = {CSCW},
  pages = {70:1--70:23},
  doi = {10.1145/3359172},
  urldate = {2024-02-05}
}

@article{blackburnHowWillYour2023,
  title = {How Will Your Relationship Be Remembered?: Virtual Relational Curation Following a Breakup},
  shorttitle = {How Will Your Relationship Be Remembered?},
  author = {Blackburn, Kate G. and LeFebvre, Leah E. and Brody, Nick},
  year = {2023},
  journal = {Information, Communication \& Society},
  volume = {0},
  number = {0},
  pages = {1--22},
  publisher = {{Routledge}},
  issn = {1369-118X},
  doi = {10.1080/1369118X.2023.2223257},
  urldate = {2023-12-08}
}

@inproceedings{gulottaDigitalArtifactsLegacy2013,
  title = {Digital Artifacts as Legacy: Exploring the Lifespan and Value of Digital Data},
  shorttitle = {Digital Artifacts as Legacy},
  booktitle = {Proceedings of the {{SIGCHI Conference}} on {{Human Factors}} in {{Computing Systems}}},
  author = {Gulotta, Rebecca and Odom, William and Forlizzi, Jodi and Faste, Haakon},
  year = {2013},
  month = apr,
  series = {{{CHI}} '13},
  pages = {1813--1822},
  publisher = {{Association for Computing Machinery}},
  address = {{New York, NY, USA}},
  doi = {10.1145/2470654.2466240},
  urldate = {2023-12-14},
  isbn = {978-1-4503-1899-0}
}

@inproceedings{caine2016,
author = {Caine, Kelly},
title = {Local Standards for Sample Size at CHI},
year = {2016},
isbn = {9781450333627},
publisher = {Association for Computing Machinery},
address = {New York, NY, USA},
url = {https://doi.org/10.1145/2858036.2858498},
doi = {10.1145/2858036.2858498},
booktitle = {Proceedings of the 2016 CHI Conference on Human Factors in Computing Systems},
pages = {981–992},
numpages = {12},
location = {San Jose, California, USA},
series = {CHI '16}
}

@article{malterudSampleSizeQualitative2016a,
  title = {Sample {{Size}} in {{Qualitative Interview Studies}}: {{Guided}} by {{Information Power}}},
  shorttitle = {Sample {{Size}} in {{Qualitative Interview Studies}}},
  author = {Malterud, Kirsti and Siersma, Volkert Dirk and Guassora, Ann Dorrit},
  year = {2016},
  month = nov,
  journal = {Qualitative Health Research},
  volume = {26},
  number = {13},
  pages = {1753--1760},
  issn = {1049-7323},
  doi = {10.1177/1049732315617444},
  langid = {english},
  pmid = {26613970}
}

@book{braun2021TA,
author={Braun, Virginia
and Clarke, Victoria},
title={Thematic analysis: A Practical Guide},
year={2021},
publisher={Sage Publications},
address={Thousand Oaks, California, United States},
pages={57-71},
isbn={9781473953246},
}

@book{riessman1993narrative,
author={Reissman, Catherine Kohler},
title={Narrative Analysis},
year={1993},
publisher={Sage Publications},
address={Newbury Park, California, United States},
isbn={9780803947542},
}

@article{apostolouWhyPeopleMake2021,
  title = {Why People Make Friends: {{The}} Nature of Friendship},
  shorttitle = {Why People Make Friends},
  author = {Apostolou, Menelaos and Keramari, Despoina and Kagialis, Antonios and Sullman, Mark},
  year = {2021},
  journal = {Personal Relationships},
  volume = {28},
  number = {1},
  pages = {4--18},
  issn = {1475-6811},
  doi = {10.1111/pere.12352},
  urldate = {2024-06-02},
  copyright = {{\copyright} 2020 International Association for Relationship Research},
  langid = {english}
}

@article{carberyFriendshipNeedFulfillment1998,
  title = {Friendship and {{Need Fulfillment During Three Phases}} of {{Young Adulthood}}},
  author = {Carbery, Julie and Buhrmester, Duane},
  year = {1998},
  month = jun,
  journal = {Journal of Social and Personal Relationships},
  volume = {15},
  number = {3},
  pages = {393--409},
  publisher = {SAGE Publications Ltd},
  issn = {0265-4075},
  doi = {10.1177/0265407598153005},
  urldate = {2024-06-02},
  langid = {english}
}

@article{clearyFriendshipMentalHealth2018,
  title = {Friendship and {{Mental Health}}},
  author = {Cleary, Michelle and Lees, David and Sayers, Jan},
  year = {2018},
  month = mar,
  journal = {Issues in Mental Health Nursing},
  publisher = {Taylor \& Francis},
  issn = {0161-2840},
  urldate = {2024-06-02},
  copyright = {{\copyright} 2018 Taylor \& Francis Group, LLC},
  langid = {english}
}

@article{demirFriendshipNeedSatisfaction2010,
  title = {Friendship, {{Need Satisfaction}} and {{Happiness}}},
  author = {Demir, Melik{\c s}ah and {\"O}zdemir, Metin},
  year = {2010},
  month = apr,
  journal = {Journal of Happiness Studies},
  volume = {11},
  number = {2},
  pages = {243--259},
  issn = {1573-7780},
  doi = {10.1007/s10902-009-9138-5},
  urldate = {2024-06-02},
  langid = {english}
}

@article{holt-lunstadWhySocialRelationships2018,
  title = {Why {{Social Relationships Are Important}} for {{Physical Health}}: {{A Systems Approach}} to {{Understanding}} and {{Modifying Risk}} and {{Protection}}},
  shorttitle = {Why {{Social Relationships Are Important}} for {{Physical Health}}},
  author = {{Holt-Lunstad}, Julianne},
  year = {2018},
  month = jan,
  journal = {Annual Review of Psychology},
  volume = {69},
  number = {Volume 69, 2018},
  pages = {437--458},
  publisher = {Annual Reviews},
  issn = {0066-4308, 1545-2085},
  doi = {10.1146/annurev-psych-122216-011902},
  urldate = {2024-06-02},
  langid = {english}
}

@article{kyleEnduringLeisureInvolvement2004,
  title = {Enduring Leisure Involvement: The Importance of Personal Relationships},
  shorttitle = {Enduring Leisure Involvement},
  author = {Kyle, {\relax Gerard} and Chick, {\relax Garry}},
  year = {2004},
  month = jul,
  journal = {Leisure Studies},
  volume = {23},
  number = {3},
  pages = {243--266},
  publisher = {Routledge},
  issn = {0261-4367},
  doi = {10.1080/0261436042000251996},
  urldate = {2024-06-02}
}

@article{lynchQUALITATIVESTUDYQUALITY2008,
  title = {A Qualitative Study of Quality of Life After Stroke: The Importance of Social Relationships},
  shorttitle = {A {{QUALITATIVE STUDY OF QUALITY OF LIFE AFTER STROKE}}},
  author = {Lynch, Elizabeth B. and Butt, Zeeshan and Heinemann, Allen and Victorson, David and Nowinski, Cindy J. and Perez, Lori and Cella, David},
  year = {2008},
  month = jul,
  journal = {Journal of rehabilitation medicine : official journal of the UEMS European Board of Physical and Rehabilitation Medicine},
  volume = {40},
  number = {7},
  pages = {10.2340/16501977-0203},
  issn = {1650-1977},
  doi = {10.2340/16501977-0203},
  urldate = {2024-06-02},
  pmcid = {PMC3869390},
  pmid = {18758667}
}

@incollection{siasFriendshipSocialSupport2007,
  title = {Friendship, {{Social Support}}, and {{Health}}},
  booktitle = {Low-{{Cost Approaches}} to {{Promote Physical}} and {{Mental Health}}: {{Theory}}, {{Research}}, and {{Practice}}},
  author = {Sias, Patricia M. and Bartoo, Heidi},
  editor = {L'Abate, Luciano},
  year = {2007},
  pages = {455--472},
  publisher = {Springer},
  address = {New York, NY},
  doi = {10.1007/0-387-36899-X_23},
  urldate = {2024-06-02},
  isbn = {978-0-387-36899-3},
  langid = {english}
}

@article{gomez-lopezWellBeingRomanticRelationships2019,
  title = {Well-{{Being}} and {{Romantic Relationships}}: {{A Systematic Review}} in {{Adolescence}} and {{Emerging Adulthood}}},
  shorttitle = {Well-{{Being}} and {{Romantic Relationships}}},
  author = {{G{\'o}mez-L{\'o}pez}, Mercedes and Viejo, Carmen and {Ortega-Ruiz}, Rosario},
  year = {2019},
  month = jan,
  journal = {International Journal of Environmental Research and Public Health},
  volume = {16},
  number = {13},
  pages = {2415},
  publisher = {Multidisciplinary Digital Publishing Institute},
  issn = {1660-4601},
  doi = {10.3390/ijerph16132415},
  urldate = {2024-06-02},
  copyright = {http://creativecommons.org/licenses/by/3.0/},
  langid = {english}
}

@article{leNeedFulfillmentEmotional2001,
  title = {Need {{Fulfillment}} and {{Emotional Experience}} in {{Interdependent Romantic Relationships}}},
  author = {Le, Benjamin and Agnew, Christopher R.},
  year = {2001},
  month = jun,
  journal = {Journal of Social and Personal Relationships},
  volume = {18},
  number = {3},
  pages = {423--440},
  publisher = {SAGE Publications Ltd},
  issn = {0265-4075},
  doi = {10.1177/0265407501183007},
  urldate = {2024-06-02},
  langid = {english}
}

@inproceedings{gashi2016unfriending,
  title={Unfriending, hiding and blocking on facebook},
  author={Gashi, Liridona and Knautz, Kathrin},
  booktitle={3rd European Conference on Social Media Research},
  pages={513--520},
  year={2016}
}

@article{overallHelpingEachOther2010,
  title = {Helping {{Each Other Grow}}: {{Romantic Partner Support}}, {{Self-Improvement}}, and {{Relationship Quality}}},
  shorttitle = {Helping {{Each Other Grow}}},
  author = {Overall, Nickola C. and Fletcher, Garth J. O. and Simpson, Jeffry A.},
  year = {2010},
  month = nov,
  journal = {Personality and Social Psychology Bulletin},
  volume = {36},
  number = {11},
  pages = {1496--1513},
  publisher = {SAGE Publications Inc},
  issn = {0146-1672},
  doi = {10.1177/0146167210383045},
  urldate = {2024-06-02},
  langid = {english}
}

@article{apostolouThisHasEnd2023,
  title = {This Has to End: {{An}} Explorative Analysis of the Strategies People Use in Order to Terminate an Undesirable Friendship},
  shorttitle = {This Has to End},
  author = {Apostolou, Menelaos},
  year = {2023},
  month = jul,
  journal = {Personality and Individual Differences},
  volume = {209},
  pages = {112211},
  issn = {0191-8869},
  doi = {10.1016/j.paid.2023.112211},
  urldate = {2024-06-02}
}

@article{bowkerWhenBestFriendships2024,
  title = {When Best Friendships End: Young Adolescents' Responses to Hypothetical Best Friendship Dissolution and Associations with Real-Life Friendship Outcomes},
  shorttitle = {When Best Friendships End},
  author = {Bowker, Julie C. and Weingarten, Jenna P. and Etkin, Rebecca G. and Dirks, Melanie A.},
  year = {2024},
  month = mar,
  journal = {Frontiers in Developmental Psychology},
  volume = {2},
  publisher = {Frontiers},
  issn = {2813-7779},
  doi = {10.3389/fdpys.2024.1369085},
  urldate = {2024-06-02},
  langid = {english}
}

@article{flanneryBreakingFriendHard2021,
  title = {Breaking {{Up}} ({{With}} a {{Friend}}) {{Is Hard}} to {{Do}}: {{An Examination}} of {{Friendship Dissolution Among Early Adolescents}}},
  shorttitle = {Breaking {{Up}} ({{With}} a {{Friend}}) {{Is Hard}} to {{Do}}},
  author = {Flannery, Kaitlin M. and Smith, Rhiannon L.},
  year = {2021},
  month = nov,
  journal = {The Journal of Early Adolescence},
  volume = {41},
  number = {9},
  pages = {1368--1393},
  publisher = {SAGE Publications Inc},
  issn = {0272-4316},
  doi = {10.1177/02724316211002266},
  urldate = {2024-06-02},
  langid = {english}
}

@article{healyWeReJust2015,
  title = {`{{We}}'re Just Not Friends Anymore': Self-Knowledge and Friendship Endings},
  shorttitle = {`{{We}}'re Just Not Friends Anymore'},
  author = {Healy, Mary},
  year = {2015},
  month = may,
  journal = {Ethics and Education},
  publisher = {Routledge},
  issn = {1744-9642},
  urldate = {2024-06-02},
  copyright = {{\copyright} 2015 Taylor \& Francis},
  langid = {english}
}

@article{roseHowFriendshipsEnd1984,
  title = {How {{Friendships End}}: {{Patterns}} among {{Young Adults}}},
  shorttitle = {How {{Friendships End}}},
  author = {Rose, Suzanna M.},
  year = {1984},
  month = sep,
  journal = {Journal of Social and Personal Relationships},
  volume = {1},
  number = {3},
  pages = {267--277},
  publisher = {SAGE Publications Ltd},
  issn = {0265-4075},
  doi = {10.1177/0265407584013001},
  urldate = {2024-06-02},
  langid = {english}
}

@article{roseKeepingEndingCasual1986,
  title = {Keeping and {{Ending Casual}}, {{Close}} and {{Best Friendships}}},
  author = {Rose, Suzanna and Serafica, Felicisima C.},
  year = {1986},
  month = sep,
  journal = {Journal of Social and Personal Relationships},
  volume = {3},
  number = {3},
  pages = {275--288},
  publisher = {SAGE Publications Ltd},
  issn = {0265-4075},
  doi = {10.1177/0265407586033002},
  urldate = {2024-06-02},
  langid = {english}
}

@article{jhaverOnlineHarassmentContent2018,
  title = {Online {{Harassment}} and {{Content Moderation}}: {{The Case}} of {{Blocklists}}},
  shorttitle = {Online {{Harassment}} and {{Content Moderation}}},
  author = {Jhaver, Shagun and Ghoshal, Sucheta and Bruckman, Amy and Gilbert, Eric},
  year = {2018},
  month = mar,
  journal = {ACM Transactions on Computer-Human Interaction},
  volume = {25},
  number = {2},
  pages = {12:1--12:33},
  issn = {1073-0516},
  doi = {10.1145/3185593},
  urldate = {2024-04-29}
}

@article{bowkerExploratoryStudyBest2023,
  title = {Exploratory Study of Best Friendship Dissolution Characteristics and Psychological Difficulties during Early Adolescence},
  author = {Bowker, Julie C. and White, Hope I. and Weingarten, Jenna P.},
  year = {2023},
  journal = {Infant and Child Development},
  volume = {32},
  number = {4},
  pages = {e2428},
  issn = {1522-7219},
  doi = {10.1002/icd.2428},
  urldate = {2024-06-03},
  copyright = {{\copyright} 2023 John Wiley \& Sons Ltd.},
  langid = {english}
}

@article{campaioliDoubleBlueTicks2022,
  title = {Double Blue Ticks: {{Reframing}} Ghosting as Ostracism through an Abductive Study on Affordances},
  shorttitle = {Double Blue Ticks},
  author = {Campaioli, Giulia and Testoni, Ines and Zamperini, Adriano},
  year = {2022},
  month = nov,
  journal = {Cyberpsychology: Journal of Psychosocial Research on Cyberspace},
  volume = {16},
  number = {5},
  issn = {1802-7962},
  doi = {10.5817/CP2022-5-10},
  urldate = {2024-04-29},
  copyright = {Copyright {\copyright} 2022 Giulia Campaioli, Ines Testoni, Adriano Zamperini},
  langid = {english}
}

@article{collinsUnwantedUnfollowedDefining2023,
  title = {Unwanted and Unfollowed: {{Defining}} Ghosting and the Role of Social Media Unfollowing},
  shorttitle = {Unwanted and Unfollowed},
  author = {Collins, Tara J. and Thomas, Angela-Faith and Harris, Emma},
  year = {2023},
  journal = {Personal Relationships},
  volume = {30},
  number = {3},
  pages = {939--959},
  issn = {1475-6811},
  doi = {10.1111/pere.12492},
  urldate = {2024-04-29},
  copyright = {{\copyright} 2023 International Association for Relationship Research.},
  langid = {english}
}

@article{freedmanEmotionalExperiencesGhosting2024,
  title = {Emotional Experiences of Ghosting},
  author = {Freedman, Gili and Powell, Darcey N. and Le, Benjamin and Williams, Kipling D.},
  year = {2024},
  month = may,
  journal = {The Journal of Social Psychology},
  volume = {164},
  number = {3},
  pages = {367--386},
  publisher = {Routledge},
  issn = {0022-4545},
  doi = {10.1080/00224545.2022.2081528},
  urldate = {2024-06-03},
  pmid = {35621208}
}

@article{kayEmpiricalAccessibleDefinition2022,
  title = {An Empirical, Accessible Definition of ``Ghosting'' as a Relationship Dissolution Method},
  author = {Kay, Caitlyn and Courtice, Erin Leigh},
  year = {2022},
  journal = {Personal Relationships},
  volume = {29},
  number = {2},
  pages = {386--411},
  issn = {1475-6811},
  doi = {10.1111/pere.12423},
  urldate = {2024-06-03},
  copyright = {{\copyright} 2022 International Association for Relationship Research.},
  langid = {english}
}

@article{pancaniGhostingOrbitingAnalysis2021a,
  title = {Ghosting and Orbiting: {{An}} Analysis of Victims' Experiences},
  shorttitle = {Ghosting and Orbiting},
  author = {Pancani, Luca and Mazzoni, Davide and Aureli, Nicolas and Riva, Paolo},
  year = {2021},
  month = jul,
  journal = {Journal of Social and Personal Relationships},
  volume = {38},
  number = {7},
  pages = {1987--2007},
  publisher = {SAGE Publications Ltd},
  issn = {0265-4075},
  doi = {10.1177/02654075211000417},
  urldate = {2024-06-03},
  langid = {english}
}

@article{pancaniRelationshipDissolutionStrategies2022a,
  title = {Relationship Dissolution Strategies: {{Comparing}} the Psychological Consequences of Ghosting, Orbiting, and Rejection},
  shorttitle = {Relationship Dissolution Strategies},
  author = {Pancani, Luca and Aureli, Nicolas and Riva, Paolo},
  year = {2022},
  month = apr,
  journal = {Cyberpsychology: Journal of Psychosocial Research on Cyberspace},
  volume = {16},
  number = {2},
  issn = {1802-7962},
  doi = {10.5817/CP2022-2-9},
  urldate = {2024-06-03},
  copyright = {Copyright {\copyright} 2022 Luca Pancani, Nicolas Aureli, and Paolo Riva},
  langid = {english}
}

@article{parkGhostingSocialRejection2024,
  title = {Ghosting: {{Social}} Rejection without Explanation, but Not without Care},
  shorttitle = {Ghosting},
  author = {Park, YeJin and Klein, Nadav},
  year = {2024},
  journal = {Journal of Experimental Psychology: General},
  pages = {No Pagination Specified-No Pagination Specified},
  publisher = {American Psychological Association},
  address = {US},
  issn = {1939-2222},
  doi = {10.1037/xge0001590}
}

@article{thomasDisappearingAgeHypervisibility2021,
  title = {Disappearing in the Age of Hypervisibility: {{Definition}}, Context, and Perceived Psychological Consequences of Social Media Ghosting},
  shorttitle = {Disappearing in the Age of Hypervisibility},
  author = {Thomas, Jhanelle Oneika and Dubar, Royette Tavernier},
  year = {2021},
  journal = {Psychology of Popular Media},
  volume = {10},
  number = {3},
  pages = {291--302},
  publisher = {Educational Publishing Foundation},
  address = {US},
  issn = {2689-6575},
  doi = {10.1037/ppm0000343}
}

@article{amichai-hamburgerFriendshipOldConcept2013,
  title = {Friendship: {{An}} Old Concept with a New Meaning?},
  shorttitle = {Friendship},
  author = {{Amichai-Hamburger}, Yair and Kingsbury, Mila and Schneider, Barry H.},
  year = {2013},
  month = jan,
  journal = {Computers in Human Behavior},
  series = {Including {{Special Section Youth}}, {{Internet}}, and {{Wellbeing}}},
  volume = {29},
  number = {1},
  pages = {33--39},
  issn = {0747-5632},
  doi = {10.1016/j.chb.2012.05.025},
  urldate = {2024-06-03}
}

@article{wuWhenSilenceSpeaks2023,
  title = {When Silence Speaks Louder than Words: {{Exploring}} the Experiences and Attitudes of Ghosters},
  shorttitle = {When Silence Speaks Louder than Words},
  author = {Wu, Karen and Bamishigbin, Olajide},
  year = {2023},
  journal = {Personal Relationships},
  volume = {30},
  number = {4},
  pages = {1358--1382},
  issn = {1475-6811},
  doi = {10.1111/pere.12518},
  urldate = {2024-04-29},
  copyright = {{\copyright} 2023 The Authors. Personal Relationships published by Wiley Periodicals LLC on behalf of International Association for Relationship Research.},
  langid = {english}
}

@article{bonettiRelationshipLonelinessSocial2010,
  title = {The {{Relationship}} of {{Loneliness}} and {{Social Anxiety}} with {{Children}}'s and {{Adolescents}}' {{Online Communication}}},
  author = {Bonetti, Luigi and Campbell, Marilyn Anne and Gilmore, Linda},
  year = {2010},
  month = jun,
  journal = {Cyberpsychology, Behavior, and Social Networking},
  volume = {13},
  number = {3},
  pages = {279--285},
  publisher = {Mary Ann Liebert, Inc., publishers},
  issn = {2152-2715},
  doi = {10.1089/cyber.2009.0215},
  urldate = {2024-06-03}
}

@article{brignalliiiImpactInternetCommunications2005,
  title = {The {{Impact}} of {{Internet Communications}} on {{Social Interaction}}},
  author = {Brignall III, Thomas Wells and Van Valey, Thomas},
  year = {2005},
  month = may,
  journal = {Sociological Spectrum},
  volume = {25},
  number = {3},
  pages = {335--348},
  publisher = {Routledge},
  issn = {0273-2173},
  doi = {10.1080/02732170590925882},
  urldate = {2024-06-03}
}

@article{chanComparisonOfflineOnline2004,
  title = {A {{Comparison}} of {{Offline}} and {{Online Friendship Qualities}} at {{Different Stages}}                of {{Relationship Development}}},
  author = {Chan, Darius K.-S. and Cheng, Grand H.-L.},
  year = {2004},
  month = jun,
  journal = {Journal of Social and Personal Relationships},
  volume = {21},
  number = {3},
  pages = {305--320},
  publisher = {SAGE Publications Ltd},
  issn = {0265-4075},
  doi = {10.1177/0265407504042834},
  urldate = {2024-06-03},
  langid = {english}
}

@article{chungSocialInteractionOnline2013,
  title = {Social Interaction in Online Support Groups: {{Preference}} for Online Social Interaction over Offline Social Interaction},
  shorttitle = {Social Interaction in Online Support Groups},
  author = {Chung, Jae Eun},
  year = {2013},
  month = jul,
  journal = {Computers in Human Behavior},
  volume = {29},
  number = {4},
  pages = {1408--1414},
  issn = {0747-5632},
  doi = {10.1016/j.chb.2013.01.019},
  urldate = {2024-06-03}
}

@article{davisFriendshipAdolescentsExperiences2012,
  title = {Friendship 2.0: {{Adolescents}}' Experiences of Belonging and Self-Disclosure Online},
  shorttitle = {Friendship 2.0},
  author = {Davis, Katie},
  year = {2012},
  month = dec,
  journal = {Journal of Adolescence},
  series = {The {{Intersection}} of {{Identity Development Processes}} and {{Peer Relationship Experiences}}},
  volume = {35},
  number = {6},
  pages = {1527--1536},
  issn = {0140-1971},
  doi = {10.1016/j.adolescence.2012.02.013},
  urldate = {2024-06-03}
}

@inproceedings{golderRhythmsSocialInteraction2007,
  title = {Rhythms of {{Social Interaction}}: {{Messaging Within}} a {{Massive Online Network}}},
  shorttitle = {Rhythms of {{Social Interaction}}},
  booktitle = {Communities and {{Technologies}} 2007},
  author = {Golder, Scott A. and Wilkinson, Dennis M. and Huberman, Bernardo A.},
  editor = {Steinfield, Charles and Pentland, Brian T. and Ackerman, Mark and Contractor, Noshir},
  year = {2007},
  pages = {41--66},
  publisher = {Springer},
  address = {London},
  doi = {10.1007/978-1-84628-905-7_3},
  isbn = {978-1-84628-905-7},
  langid = {english}
}

@article{hoodLonelinessOnlineFriendships2018,
  title = {Loneliness and Online Friendships in Emerging Adults},
  author = {Hood, Michelle and Creed, Peter A. and Mills, Bianca J.},
  year = {2018},
  month = oct,
  journal = {Personality and Individual Differences},
  series = {Examining {{Personality}} and {{Individual Differences}} in {{Cyberspace}}},
  volume = {133},
  pages = {96--102},
  issn = {0191-8869},
  doi = {10.1016/j.paid.2017.03.045},
  urldate = {2024-06-03}
}

@article{mittmannTikTokMyLife2022,
  title = {``{{TikTok Is My Life}} and {{Snapchat Is My Ventricle}}'': {{A Mixed-Methods Study}} on the {{Role}} of {{Online Communication Tools}} for {{Friendships}} in {{Early Adolescents}}},
  shorttitle = {``{{TikTok Is My Life}} and {{Snapchat Is My Ventricle}}''},
  author = {Mittmann, Gloria and Woodcock, Kate and D{\"o}rfler, Sylvia and Krammer, Ina and Pollak, Isabella and Schrank, Beate},
  year = {2022},
  month = feb,
  journal = {The Journal of Early Adolescence},
  volume = {42},
  number = {2},
  pages = {172--203},
  publisher = {SAGE Publications Inc},
  issn = {0272-4316},
  doi = {10.1177/02724316211020368},
  urldate = {2024-04-29},
  langid = {english}
}

@article{nesiTransformationAdolescentPeer2018,
  title = {Transformation of {{Adolescent Peer Relations}} in the {{Social Media Context}}: {{Part}} 1---{{A Theoretical Framework}} and {{Application}} to {{Dyadic Peer Relationships}}},
  shorttitle = {Transformation of {{Adolescent Peer Relations}} in the {{Social Media Context}}},
  author = {Nesi, Jacqueline and {Choukas-Bradley}, Sophia and Prinstein, Mitchell J.},
  year = {2018},
  month = sep,
  journal = {Clinical Child and Family Psychology Review},
  volume = {21},
  number = {3},
  pages = {267--294},
  issn = {1573-2827},
  doi = {10.1007/s10567-018-0261-x},
  urldate = {2024-06-03},
  langid = {english}
}

@article{reichFriendingIMingHanging2012,
  title = {Friending, {{IMing}}, and Hanging out Face-to-Face: {{Overlap}} in Adolescents' Online and Offline Social Networks},
  shorttitle = {Friending, {{IMing}}, and Hanging out Face-to-Face},
  author = {Reich, Stephanie M. and Subrahmanyam, Kaveri and Espinoza, Guadalupe},
  year = {2012},
  journal = {Developmental Psychology},
  volume = {48},
  number = {2},
  pages = {356--368},
  publisher = {American Psychological Association},
  address = {US},
  issn = {1939-0599},
  doi = {10.1037/a0026980}
}

@article{tangDevelopmentOnlineFriendship2010,
  title = {Development of {{Online Friendship}} in {{Different Social Spaces}}: {{A}} Case Study},
  shorttitle = {Development of {{Online Friendship}} in {{Different Social Spaces}}},
  author = {Tang, Lijun},
  year = {2010},
  month = jun,
  journal = {Information, Communication \& Society},
  volume = {13},
  number = {4},
  pages = {615--633},
  publisher = {Routledge},
  issn = {1369-118X},
  doi = {10.1080/13691180902998639},
  urldate = {2024-06-03}
}

@article{betzAmbiguousLossFamily2006a,
  title = {Ambiguous {{Loss}} and the {{Family Grieving Process}}},
  author = {Betz, Gabrielle and Thorngren, Jill M.},
  year = {2006},
  month = oct,
  journal = {The Family Journal},
  volume = {14},
  number = {4},
  pages = {359--365},
  publisher = {SAGE Publications Inc},
  issn = {1066-4807},
  doi = {10.1177/1066480706290052},
  urldate = {2024-08-05},
  langid = {english}
}

@article{bossAmbiguousLossComplicated2014,
  title = {Ambiguous Loss: A Complicated Type of Grief When Loved Ones Disappear},
  shorttitle = {Ambiguous Loss},
  author = {Boss, Pauline and Yeats, Janet R},
  year = {2014},
  month = may,
  journal = {Bereavement Care},
  volume = {33},
  number = {2},
  pages = {63--69},
  publisher = {Routledge},
  issn = {0268-2621},
  doi = {10.1080/02682621.2014.933573},
  urldate = {2024-08-05}
}

@article{bossAmbiguousLossTheory2007,
  title = {Ambiguous {{Loss Theory}}: {{Challenges}} for {{Scholars}} and {{Practitioners}}},
  shorttitle = {Ambiguous {{Loss Theory}}},
  author = {Boss, Pauline},
  year = {2007},
  journal = {Family Relations},
  volume = {56},
  number = {2},
  eprint = {4541653},
  eprinttype = {jstor},
  pages = {105--110},
  publisher = {[National Council on Family Relations, Wiley]},
  issn = {0197-6664},
  urldate = {2024-08-05}
}

@article{bossTraumaComplicatedGrief2010,
  title = {The {{Trauma}} and {{Complicated Grief}} of {{Ambiguous Loss}}},
  author = {Boss, Pauline},
  year = {2010},
  month = apr,
  journal = {Pastoral Psychology},
  volume = {59},
  number = {2},
  pages = {137--145},
  issn = {1573-6679},
  doi = {10.1007/s11089-009-0264-0},
  urldate = {2024-08-05},
  langid = {english}
}

@article{wieseDynamicalSocialPsychology2010,
  title = {Dynamical {{Social Psychology}}: {{Complexity}} and {{Coherence}} in {{Human Experience}}},
  shorttitle = {Dynamical {{Social Psychology}}},
  author = {Wiese, Susan L. and Vallacher, Robin R. and Strawinska, Urszula},
  year = {2010},
  journal = {Social and Personality Psychology Compass},
  volume = {4},
  number = {11},
  pages = {1018--1030},
  issn = {1751-9004},
  doi = {10.1111/j.1751-9004.2010.00319.x},
  urldate = {2024-08-09},
  copyright = {{\copyright} 2010 The Authors. Social and Personality Psychology Compass {\copyright} 2010 Blackwell Publishing Ltd},
  langid = {english}
}

@article{peterson2006people,
  title={People are complex and the world is messy: A behavior-based approach to executive coaching},
  author={Peterson, David B},
  journal={Evidence based coaching handbook: Putting best practices to work for your clients},
  pages={51--76},
  year={2006},
  publisher={John Wiley \& Sons Hoboken, NJ}
}

@article{crawfordTimeHealsAll2023,
  title = {Time Heals All Wounds? {{Na{\"i}ve}} Theories about the Fading of Affect Associated with Autobiographical Events},
  shorttitle = {Time Heals All Wounds?},
  author = {Crawford, Matthew T. and Marsh, Claire},
  year = {2023},
  month = nov,
  journal = {Memory \& Cognition},
  volume = {51},
  number = {8},
  pages = {1715--1728},
  issn = {1532-5946},
  doi = {10.3758/s13421-023-01426-2},
  urldate = {2024-08-15},
  langid = {english}
}

@article{walkerFadingAffectBias2009,
  title = {The {{Fading}} Affect Bias: {{But}} What the Hell Is It For?},
  shorttitle = {The {{Fading}} Affect Bias},
  author = {Walker, W. Richard and Skowronski, John J.},
  year = {2009},
  journal = {Applied Cognitive Psychology},
  volume = {23},
  number = {8},
  pages = {1122--1136},
  issn = {1099-0720},
  doi = {10.1002/acp.1614},
  urldate = {2024-08-15},
  langid = {english}
}

@article{karremansBackCaringBeing2004,
  title = {Back to Caring after Being Hurt: The Role of Forgiveness},
  shorttitle = {Back to Caring after Being Hurt},
  author = {Karremans, Johan C. and Van Lange, Paul A. M.},
  year = {2004},
  journal = {European Journal of Social Psychology},
  volume = {34},
  number = {2},
  pages = {207--227},
  issn = {1099-0992},
  doi = {10.1002/ejsp.192},
  urldate = {2024-08-08},
  copyright = {Copyright {\copyright} 2004 John Wiley \& Sons, Ltd.},
  langid = {english}
}

@article{pasupathiWhenHurtOthers2019,
  title = {When {{I}} Hurt Others, and When {{I}} Get Hurt: {{Integrating}} Victim and Perpetrator Experiences of Harm into a Sense of Moral Agency},
  shorttitle = {When {{I}} Hurt Others, and When {{I}} Get Hurt},
  author = {Pasupathi, Monisha and Wainryb, Cecilia},
  year = {2019},
  journal = {Social Development},
  volume = {28},
  number = {4},
  pages = {820--834},
  issn = {1467-9507},
  doi = {10.1111/sode.12334},
  urldate = {2024-08-08},
  copyright = {{\copyright} 2018 John Wiley \& Sons Ltd},
  langid = {english}
}

@article{vangelistiReactionsMessagesThat1998,
  title = {Reactions to Messages That Hurt: {{The}} Influence of Relational Contexts},
  shorttitle = {Reactions to Messages That Hurt},
  author = {Vangelisti, Anita L. and Crumley, Linda P.},
  year = {1998},
  month = sep,
  journal = {Communication Monographs},
  volume = {65},
  number = {3},
  pages = {173--196},
  publisher = {Routledge},
  issn = {0363-7751},
  doi = {10.1080/03637759809376447},
  urldate = {2024-08-08}
}

@inproceedings{vasalouApplicationForgivenessSocial2009,
  title = {The Application of Forgiveness in Social System Design},
  booktitle = {Proceedings of the {{SIGCHI Conference}} on {{Human Factors}} in {{Computing Systems}}},
  author = {Vasalou, Asimina and Riegelsberger, Jens and Joinson, Adam},
  year = {2009},
  month = apr,
  series = {{{CHI}} '09},
  pages = {225--228},
  publisher = {Association for Computing Machinery},
  address = {New York, NY, USA},
  doi = {10.1145/1518701.1518738},
  urldate = {2024-08-07},
  isbn = {978-1-60558-246-7}
}

@article{vasalouPraiseForgivenessWays2008,
  title = {In Praise of Forgiveness: {{Ways}} for Repairing Trust Breakdowns in One-off Online Interactions},
  shorttitle = {In Praise of Forgiveness},
  author = {Vasalou, Asimina and Hopfensitz, Astrid and Pitt, Jeremy V.},
  year = {2008},
  month = jun,
  journal = {International Journal of Human-Computer Studies},
  volume = {66},
  number = {6},
  pages = {466--480},
  issn = {1071-5819},
  doi = {10.1016/j.ijhcs.2008.02.001},
  urldate = {2024-08-08}
}

@article{kannabiras2021sorry,
author = {Kannabiran, Gopinaath},
title = {I am sorry!},
year = {2021},
issue_date = {September - October 2021},
publisher = {Association for Computing Machinery},
address = {New York, NY, USA},
volume = {28},
number = {5},
issn = {1072-5520},
url = {https://doi.org/10.1145/3477089},
doi = {10.1145/3477089},
journal = {Interactions},
month = {aug},
pages = {14–15},
numpages = {2}
}

@article{lefebvreGhostedNavigatingStrategies2020,
  title = {Ghosted?: {{Navigating}} Strategies for Reducing Uncertainty and Implications Surrounding Ambiguous Loss},
  shorttitle = {Ghosted?},
  author = {LeFebvre, Leah E. and Fan, Xiaoti},
  year = {2020},
  journal = {Personal Relationships},
  volume = {27},
  number = {2},
  pages = {433--459},
  issn = {1475-6811},
  doi = {10.1111/pere.12322},
  urldate = {2024-08-15},
  copyright = {{\copyright} 2020 IARR},
  langid = {english}
}

@incollection{wilsonfadijiExploringMeaningMakingUniversity2022,
  title = {Exploring {{Meaning-Making Among University Students}} in {{South Africa During}} the {{COVID-19 Lockdown}}},
  booktitle = {Emerging {{Adulthood}} in the {{COVID-19 Pandemic}} and {{Other Crises}}: {{Individual}} and {{Relational Resources}}},
  author = {Wilson Fadiji, Angelina and Chigeza, Shingairai and Shoko, Placidia},
  editor = {Leontopoulou, Sophie and Delle Fave, Antonella},
  year = {2022},
  series = {Cross-{{Cultural Advancements}} in {{Positive Psychology}}},
  pages = {97--115},
  publisher = {{Springer International Publishing}},
  address = {{Cham}},
  doi = {10.1007/978-3-031-22288-7_7},
  urldate = {2024-02-21},
  isbn = {978-3-031-22288-7},
  langid = {english}
}

@article{todorovaWhatThoughtWas2021,
  title = {``{{What I}} Thought Was so Important Isn't Really That Important'': International Perspectives on Making Meaning during the First Wave of the {{COVID-19}} Pandemic},
  shorttitle = {``{{What I}} Thought Was so Important Isn't Really That Important''},
  author = {Todorova, Irina and Albers, Liesemarie and Aronson, Nicole and Baban, Adriana and Benyamini, Yael and Cipolletta, Sabrina and {del Rio Carral}, Maria and Dimitrova, Elitsa and Dudley, Claire and Guzzardo, Mariana and Hammoud, Razan and Fadil Azim, Darlina Hani and Hilverda, Femke and Huang, Qi and John, Liji and Kaneva, Michaela and Khan, Sanjida and Kostova, Zlatina and Kotzeva, Tatyana and Fathima, M.A. and Anto, Milu Maria and Michoud, Chlo{\'e} and Awal Miah, Mohammad Abdul and Mohr, Julia and Morgan, Karen and Nastase, Elena Simona and Neter, Efrat and Panayotova, Yulia and Patel, Hemali and Pillai, Dhanya and Polidoro Lima, Manuela and Qin, Desiree Baolian and Salewski, Christel and Sankar, K. Anu and Shao, Sabrina and Suresh, Jeevanisha and Todorova, Ralitsa and Tomaino, Silvia Caterina Maria and Vollmann, Manja and Winter, David and Xie, Mingjun and Xuan Ning, Sam and Zlatarska, Asya},
  year = {2021},
  month = jan,
  journal = {Health Psychology and Behavioral Medicine},
  volume = {9},
  number = {1},
  pages = {830--857},
  publisher = {{Routledge}},
  issn = {null},
  doi = {10.1080/21642850.2021.1981909},
  urldate = {2024-02-21},
  pmid = {34650834}
}

@article{morganWhyMeaningmakingMatters2020,
  title = {Why Meaning-Making Matters: The Case of the {{UK Government}}'s {{COVID-19}} Response},
  shorttitle = {Why Meaning-Making Matters},
  author = {Morgan, Marcus},
  year = {2020},
  month = dec,
  journal = {American Journal of Cultural Sociology},
  volume = {8},
  number = {3},
  pages = {270--323},
  issn = {2049-7121},
  doi = {10.1057/s41290-020-00121-y},
  urldate = {2024-02-21},
  langid = {english}
}

@article{castiglioniFosteringReconstructionMeaning2020,
  title = {Fostering the {{Reconstruction}} of {{Meaning Among}} the {{General Population During}} the {{COVID-19 Pandemic}}},
  author = {Castiglioni, Marco and Gaj, Nicolo'},
  year = {2020},
  journal = {Frontiers in Psychology},
  volume = {11},
  issn = {1664-1078},
  urldate = {2024-02-21}
}

@article{corina2016grief,
author = {Sas, Corina and Whittaker, Steve and Zimmerman, John},
title = {Design for Rituals of Letting Go: An Embodiment Perspective on Disposal Practices Informed by Grief Therapy},
year = {2016},
issue_date = {September 2016},
publisher = {Association for Computing Machinery},
address = {New York, NY, USA},
volume = {23},
number = {4},
issn = {1073-0516},
url = {https://doi.org/10.1145/2926714},
doi = {10.1145/2926714},
journal = {ACM Trans. Comput.-Hum. Interact.},
month = {aug},
articleno = {21},
numpages = {37}
}

@inproceedings{ravn2024materializing,
author = {Ravn, Nicklas and Striib, Nadja and Fritsch, Jonas},
title = {Materialising affective experiences: Designing for personal domestic grief practices},
year = {2024},
isbn = {9798400705830},
publisher = {Association for Computing Machinery},
address = {New York, NY, USA},
url = {https://doi.org/10.1145/3643834.3660753},
doi = {10.1145/3643834.3660753},
booktitle = {Proceedings of the 2024 ACM Designing Interactive Systems Conference},
pages = {685–698},
numpages = {14},
location = {Copenhagen, Denmark},
series = {DIS '24}
}

@article{biolcatiCyberDatingAbuse2021,
  title = {Cyber Dating Abuse and Ghosting Behaviours: Personality and Gender Roles in Romantic Relationships},
  shorttitle = {Cyber Dating Abuse and Ghosting Behaviours},
  author = {Biolcati, Roberta and Pupi, Virginia and Mancini, Giacomo},
  year = {2021},
  month = sep,
  journal = {Current Issues in Personality Psychology},
  volume = {10},
  number = {3},
  pages = {240--251},
  issn = {2353-4192},
  doi = {10.5114/cipp.2021.108289},
  urldate = {2024-08-15},
  pmcid = {PMC10535627},
  pmid = {38013819}
}

@article{navarroPsychologicalCorrelatesGhosting2020,
  title = {Psychological {{Correlates}} of {{Ghosting}} and {{Breadcrumbing Experiences}}: {{A Preliminary Study}} among {{Adults}}},
  shorttitle = {Psychological {{Correlates}} of {{Ghosting}} and {{Breadcrumbing Experiences}}},
  author = {Navarro, Ra{\'u}l and Larra{\~n}aga, Elisa and Yubero, Santiago and V{\'i}llora, Beatriz},
  year = {2020},
  month = jan,
  journal = {International Journal of Environmental Research and Public Health},
  volume = {17},
  number = {3},
  pages = {1116},
  publisher = {Multidisciplinary Digital Publishing Institute},
  issn = {1660-4601},
  doi = {10.3390/ijerph17031116},
  urldate = {2024-08-15},
  copyright = {http://creativecommons.org/licenses/by/3.0/},
  langid = {english}
}

@inproceedings{jarupreechachanNotDisturbImplication2023,
  title = {`{{Do Not Disturb}}': {{An Implication Design To Be Alerted But Less Stress}}},
  shorttitle = {`{{Do Not Disturb}}'},
  booktitle = {2023 {{International Seminar}} on {{Application}} for {{Technology}} of {{Information}} and {{Communication}} ({{iSemantic}})},
  author = {Jarupreechachan, Weerachaya and Kitchat, Kotcharat and Surasak, Thattapon},
  year = {2023},
  month = sep,
  pages = {176--181},
  doi = {10.1109/iSemantic59612.2023.10295289},
  urldate = {2024-08-15}
}

@article{sihombingPhenomenologyUsingInstagram2022,
  title = {Phenomenology {{Of Using Instagram Close Friend Features For Self Disclosure Improvement}}},
  author = {Sihombing, Lambok Hermanto and Aninda, Maria Paskalia},
  year = {2022},
  month = jun,
  journal = {Professional: Jurnal Komunikasi dan Administrasi Publik},
  volume = {9},
  number = {1},
  pages = {29--34},
  issn = {2722-371X},
  doi = {10.37676/professional.v9i1.2282},
  urldate = {2024-08-15},
  copyright = {Copyright (c) 2022 Lambok Hermanto Sihombing, Maria Paskalia Aninda},
  langid = {english}
}

@article{cillessenPredictorsDyadicFriendship2005,
  title = {Predictors of Dyadic Friendship Quality in Adolescence},
  author = {Cillessen, Antonius and Jiang, X. Lu and West, Tessa and Laszkowski, Dagmara},
  year = {2005},
  month = jan,
  journal = {International Journal of Behavioral Development},
  volume = {29},
  number = {2},
  pages = {165--172},
  publisher = {Routledge},
  issn = {0165-0254},
  doi = {10.1080/01650250444000360},
  urldate = {2024-08-09}
}

@article{oswaldFriendshipMaintenanceAnalysis2004,
  title = {Friendship {{Maintenance}}: {{An Analysis}} of {{Individual}} and {{Dyad Behaviors}}},
  shorttitle = {Friendship {{Maintenance}}},
  author = {Oswald, Debra L. and Clark, Eddie M. and Kelly, Cheryl M.},
  year = {2004},
  month = jun,
  journal = {Journal of Social and Clinical Psychology},
  volume = {23},
  number = {3},
  pages = {413--441},
  publisher = {Guilford Publications Inc.},
  issn = {0736-7236},
  doi = {10.1521/jscp.23.3.413.35460},
  urldate = {2024-08-09}
}

@article{hallFriendshipStandardsDimensions2012,
  title = {Friendship Standards: {{The}} Dimensions of Ideal Expectations},
  shorttitle = {Friendship Standards},
  author = {Hall, Jeffrey A.},
  year = {2012},
  month = nov,
  journal = {Journal of Social and Personal Relationships},
  volume = {29},
  number = {7},
  pages = {884--907},
  publisher = {SAGE Publications Ltd},
  issn = {0265-4075},
  doi = {10.1177/0265407512448274},
  urldate = {2024-08-16},
  langid = {english}
}

@article{clarkFriendshipExpectationsFriendship1993,
  title = {Friendship Expectations and Friendship Evaluations: {{Reciprocity}} and Gender Effects},
  shorttitle = {Friendship Expectations and Friendship Evaluations},
  author = {Clark, M. L. and Ayers, Marla},
  year = {1993},
  journal = {Youth \& Society},
  volume = {24},
  number = {3},
  pages = {299--313},
  publisher = {Sage Publications},
  address = {US},
  issn = {1552-8499},
  doi = {10.1177/0044118X93024003003}
}

@article{foxAgeGenderDimensions1985b,
  title = {Age and {{Gender Dimensions}} of {{Friendship}}},
  author = {Fox, Margery and Gibbs, Margaret and Auerbach, Doris},
  year = {1985},
  month = dec,
  journal = {Psychology of Women Quarterly},
  volume = {9},
  number = {4},
  pages = {489--502},
  publisher = {SAGE Publications Inc},
  issn = {0361-6843},
  doi = {10.1111/j.1471-6402.1985.tb00898.x},
  urldate = {2024-08-18},
  langid = {english}
}

@article{macevoyWhenFriendsDisappoint2012,
  title = {When {{Friends Disappoint}}: {{Boys}}' and {{Girls}}' {{Responses}} to {{Transgressions}} of {{Friendship Expectations}}},
  shorttitle = {When {{Friends Disappoint}}},
  author = {MacEvoy, Julie Paquette and Asher, Steven R.},
  year = {2012},
  journal = {Child Development},
  volume = {83},
  number = {1},
  pages = {104--119},
  issn = {1467-8624},
  doi = {10.1111/j.1467-8624.2011.01685.x},
  urldate = {2024-08-18},
  copyright = {{\copyright} 2011 The Authors. Child Development {\copyright} 2011 Society for Research in Child Development, Inc},
  langid = {english}
}

@article{burgoonInterpersonalExpectationsExpectancy1993,
  title = {Interpersonal {{Expectations}}, {{Expectancy Violations}}, and {{Emotional Communication}}},
  author = {Burgoon, Judee K.},
  year = {1993},
  month = mar,
  journal = {Journal of Language and Social Psychology},
  volume = {12},
  number = {1-2},
  pages = {30--48},
  publisher = {SAGE Publications Inc},
  issn = {0261-927X},
  doi = {10.1177/0261927X93121003},
  urldate = {2025-02-11},
  langid = {english}
}

@article{waltherComputerMediatedCommunicationImpersonal1996b,
  title = {Computer-{{Mediated Communication}}: {{Impersonal}}, {{Interpersonal}}, and {{Hyperpersonal Interaction}}},
  shorttitle = {Computer-{{Mediated Communication}}},
  author = {WALTHER, JOSEPH B.},
  year = {1996},
  month = feb,
  journal = {Communication Research},
  volume = {23},
  number = {1},
  pages = {3--43},
  publisher = {SAGE Publications Inc},
  issn = {0093-6502},
  doi = {10.1177/009365096023001001},
  urldate = {2025-02-11},
  langid = {english}
}

@article{walther2011theories,
author = {Walther, Joseph},
year = {2011},
month = {01},
pages = {443-479},
title = {Theories of computer-mediated communication and interpersonal relations},
journal = {The Handbook of Interpersonal Communication}
}

@inproceedings{kwakFragileOnlineRelationship2011,
  title = {Fragile Online Relationship: A First Look at Unfollow Dynamics in Twitter},
  shorttitle = {Fragile Online Relationship},
  booktitle = {Proceedings of the {{SIGCHI Conference}} on {{Human Factors}} in {{Computing Systems}}},
  author = {Kwak, Haewoon and Chun, Hyunwoo and Moon, Sue},
  year = {2011},
  month = may,
  series = {{{CHI}} '11},
  pages = {1091--1100},
  publisher = {Association for Computing Machinery},
  address = {New York, NY, USA},
  doi = {10.1145/1978942.1979104},
  urldate = {2025-05-05},
  isbn = {978-1-4503-0228-9}
}

@article{karr-wisniewskiNewSocialOrder2011a,
  title = {A {{New Social Order}}: {{Mechanisms}} for {{Social Network Site Boundary Regulation}}},
  shorttitle = {A {{New Social Order}}},
  author = {{Karr-Wisniewski}, Pamela and Wilson, D. and {Richter-Lipford}},
  year = {2011},
  month = aug,
  journal = {AMCIS 2011 Proceedings - All Submissions}
}

@inproceedings{wisniewskiFightingMySpace2012,
  title = {Fighting for My Space: Coping Mechanisms for Sns Boundary Regulation},
  shorttitle = {Fighting for My Space},
  booktitle = {Proceedings of the {{SIGCHI Conference}} on {{Human Factors}} in {{Computing Systems}}},
  author = {Wisniewski, Pamela and Lipford, Heather and Wilson, David},
  year = {2012},
  month = may,
  series = {{{CHI}} '12},
  pages = {609--618},
  publisher = {Association for Computing Machinery},
  address = {New York, NY, USA},
  doi = {10.1145/2207676.2207761},
  urldate = {2025-09-14},
  isbn = {978-1-4503-1015-4}
}

@article{atkins1993reflection,
  title={Reflection: a review of the literature},
  author={Atkins, Sue and Murphy, Kathy},
  journal={Journal of advanced nursing},
  volume={18},
  number={8},
  pages={1188--1192},
  year={1993},
  publisher={Wiley Online Library}
}

@article{bossMythClosure2012,
  title = {The {{Myth}} of {{Closure}}},
  author = {Boss, Pauline and Carnes, Donna},
  year = {2012},
  journal = {Family Process},
  volume = {51},
  number = {4},
  pages = {456--469},
  issn = {1545-5300},
  doi = {10.1111/famp.12005},
  urldate = {2025-09-15},
  copyright = {{\copyright} FPI, Inc.},
  langid = {english}
}

@article{curciNegativeEmotionalExperiences2013,
  title = {Negative Emotional Experiences Arouse Rumination and Affect Working Memory Capacity.},
  author = {Curci, Antonietta and Lanciano, Tiziana and Soleti, Emanuela and Rim{\'e}, Bernard},
  year = {2013},
  month = oct,
  journal = {Emotion},
  volume = {13},
  number = {5},
  pages = {867--880},
  publisher = {American Psychological Association},
  issn = {1931-1516},
  doi = {10.1037/a0032492},
  urldate = {2025-09-15},
  langid = {english}
}

@article{kirkegaardthomsenAssociationRuminationNegative2006,
  title = {The Association between Rumination and Negative Affect: {{A}} Review},
  shorttitle = {The Association between Rumination and Negative Affect},
  author = {Kirkegaard Thomsen, Dorthe},
  year = {2006},
  month = dec,
  journal = {Cognition and Emotion},
  volume = {20},
  number = {8},
  pages = {1216--1235},
  publisher = {Routledge},
  issn = {0269-9931},
  doi = {10.1080/02699930500473533},
  urldate = {2025-09-15}
}

@article{bryant2005using,
  title={Using the past to enhance the present: Boosting happiness through positive reminiscence},
  author={Bryant, Fred B. and Smart, Colette M. and King, Scott P.},
  journal={Journal of Happiness Studies},
  volume={6},
  number={3},
  pages={227--260},
  year={2005},
  doi = {10.1007/s10902-005-3889-4},
  publisher={Springer}
}

@article{lyubomirsky2005pursuing,
  title={Pursuing happiness: The architecture of sustainable change},
  author={Lyubomirsky, Sonja and Sheldon, Kennon M and Schkade, David},
  journal={Review of general psychology},
  volume={9},
  number={2},
  pages={111--131},
  year={2005},
  publisher={SAGE Publications Sage CA: Los Angeles, CA}
}

@inproceedings{slovak2017ref-practicum,
author = {Slov\'{a}k, Petr and Frauenberger, Christopher and Fitzpatrick, Geraldine},
title = {Reflective Practicum: A Framework of Sensitising Concepts to Design for Transformative Reflection},
year = {2017},
publisher = {Association for Computing Machinery},
address = {NY, USA},
doi = {10.1145/3025453.3025516},
booktitle = {Proceedings of the 2017 CHI Conference on Human Factors in Computing Systems},
pages = {2696–2707},
numpages = {12},
location = {Denver, Colorado, USA},
series = {CHI '17}
}

@article{cowanFacilitatingReflectiveJournalling2013a,
  title = {Facilitating Reflective Journalling},
  author = {Cowan, John},
  year = {2013},
  month = mar,
  journal = {Journal of Learning Development in Higher Education},
  number = {5},
  issn = {1759-667X},
  doi = {10.47408/jldhe.v0i5.154},
  urldate = {2025-08-09},
  copyright = {Copyright (c)},
  langid = {english}
}

@article{richelleStayPositiveEffects2024a,
  title = {Stay {{Positive}}: {{The Effects}} of {{Positive Affect Journaling}} on {{Emotion When Remembering COVID-19}}},
  shorttitle = {Stay {{Positive}}},
  author = {Richelle, Justine and Alea, Nicole},
  year = {2024},
  month = oct,
  journal = {Journal of Creativity in Mental Health},
  volume = {19},
  number = {4},
  pages = {529--541},
  publisher = {Routledge},
  issn = {1540-1383},
  doi = {10.1080/15401383.2023.2281547},
  urldate = {2025-09-15}
}

@article{ullrichJournalingStressfulEvents2002c,
  title = {Journaling about Stressful Events: {{Effects}} of Cognitive Processing and Emotional Expression},
  shorttitle = {Journaling about Stressful Events},
  author = {Ullrich, Philip M. and Lutgendorf, Susan K.},
  year = {2002},
  month = aug,
  journal = {Annals of Behavioral Medicine},
  volume = {24},
  number = {3},
  pages = {244--250},
  issn = {0883-6612},
  doi = {10.1207/S15324796ABM2403_10},
  urldate = {2025-08-10}
}

@incollection{walkerWritingReflection1985,
  title = {Writing and {{Reflection}}},
  booktitle = {Reflection},
  author = {Walker, David},
  year = {1985},
  publisher = {Routledge},
  isbn = {978-1-315-05905-1}
}

@inproceedings{walter2008journaling,
  title={Journaling as writing practice, reflection, and personal expression},
  author={Walter-Echols, Elizabeth},
  booktitle={CamTESOL conference on english language teaching: Selected papers},
  volume={4},
  pages={120--131},
  year={2008}
}

@inproceedings{zhouJournalAIdeEmpoweringOlder2025,
  title = {{{JournalAIde}}: {{Empowering Older Adults}} in {{Digital Journal Writing}}},
  shorttitle = {{{JournalAIde}}},
  booktitle = {Proceedings of the 2025 {{CHI Conference}} on {{Human Factors}} in {{Computing Systems}}},
  author = {Zhou, Shixu and Lin, Weiyue and Xu, Zuyu and Wei, Xiaoying and Huang, Raoyi and Ma, Xiaojuan and Fan, Mingming},
  year = {2025},
  month = apr,
  series = {{{CHI}} '25},
  pages = {1--19},
  publisher = {Association for Computing Machinery},
  address = {New York, NY, USA},
  doi = {10.1145/3706598.3713339},
  urldate = {2025-07-19},
  isbn = {979-8-4007-1394-1}
}

@article{muradovaSeeingOtherSide2021,
  title = {Seeing the {{Other Side}}? {{Perspective-Taking}} and {{Reflective Political Judgements}} in {{Interpersonal Deliberation}}},
  shorttitle = {Seeing the {{Other Side}}?},
  author = {Muradova, Lala},
  year = {2021},
  month = aug,
  journal = {Political Studies},
  volume = {69},
  number = {3},
  pages = {644--664},
  publisher = {SAGE Publications Ltd},
  issn = {0032-3217},
  doi = {10.1177/0032321720916605},
  urldate = {2025-09-15},
  langid = {english}
}

@article{southworthBridgingCriticalThinking2022,
  title = {Bridging Critical Thinking and Transformative Learning: {{The}} Role of Perspective-Taking},
  shorttitle = {Bridging Critical Thinking and Transformative Learning},
  author = {Southworth, James},
  year = {2022},
  month = mar,
  journal = {Theory and Research in Education},
  volume = {20},
  number = {1},
  pages = {44--63},
  publisher = {SAGE Publications},
  issn = {1477-8785},
  doi = {10.1177/14778785221090853},
  urldate = {2025-09-15},
  langid = {english}
}

@article{whitakerMakingArrangementsCuration2022,
  title = {Making {{Arrangements}}: {{The Curation}} of {{Grief}} in the {{Home Studio}} ({{Faire}} Les Arrangements : La Conservation Du Deuil Dans Le Studio {\`a} Domicile)},
  shorttitle = {Making {{Arrangements}}},
  author = {Whitaker, Pamela and McHugh, Christopher},
  year = {2022},
  month = jan,
  journal = {Canadian Journal of Art Therapy},
  volume = {35},
  number = {1},
  pages = {32--42},
  publisher = {Routledge},
  issn = {2690-7240},
  doi = {10.1080/26907240.2021.1976368},
  urldate = {2025-09-15}
}

@inproceedings{wagenerMoodShaper2024,
author = {Wagener, Nadine and Kiesewetter, Arne and Reicherts, Leon and Wo\'{z}niak, Pawe\l{} W. and Sch\"{o}ning, Johannes and Rogers, Yvonne and Niess, Jasmin},
title = {MoodShaper: A Virtual Reality Experience to Support Managing Negative Emotions},
year = {2024},
isbn = {9798400705830},
publisher = {Association for Computing Machinery},
address = {New York, NY, USA},
url = {https://doi.org/10.1145/3643834.3661570},
doi = {10.1145/3643834.3661570},
booktitle = {Proceedings of the 2024 ACM Designing Interactive Systems Conference},
pages = {2286–2304},
numpages = {19},
location = {Copenhagen, Denmark},
series = {DIS '24}
}

@article{lewisApologiesCloseRelationships2015,
  title = {Apologies in {{Close Relationships}}: {{A Review}} of {{Theory}} and {{Research}}},
  shorttitle = {Apologies in {{Close Relationships}}},
  author = {Lewis, Jarrett T. and Parra, Gilbert R. and Cohen, Robert},
  year = {2015},
  journal = {Journal of Family Theory \& Review},
  volume = {7},
  number = {1},
  pages = {47--61},
  issn = {1756-2589},
  doi = {10.1111/jftr.12060},
  urldate = {2025-09-15},
  copyright = {{\copyright} 2015 National Council on Family Relations},
  langid = {english}
}

@article{schumannDoesLoveMean2012,
  title = {Does Love Mean Never Having to Say You're Sorry? {{Associations}} between Relationship Satisfaction, Perceived Apology Sincerity, and Forgiveness},
  shorttitle = {Does Love Mean Never Having to Say You're Sorry?},
  author = {Schumann, Karina},
  year = {2012},
  month = nov,
  journal = {Journal of Social and Personal Relationships},
  volume = {29},
  number = {7},
  pages = {997--1010},
  publisher = {SAGE Publications Ltd},
  issn = {0265-4075},
  doi = {10.1177/0265407512448277},
  urldate = {2025-09-15},
  langid = {english}
}

@article{battistiSecondPersonAuthenticityMediating2025,
  title = {Second-{{Person Authenticity}} and the {{Mediating Role}} of {{AI}}: {{A Moral Challenge}} for {{Human-to-Human Relationships}}?},
  shorttitle = {Second-{{Person Authenticity}} and the {{Mediating Role}} of {{AI}}},
  author = {Battisti, Davide},
  year = {2025},
  month = feb,
  journal = {Philosophy \& Technology},
  volume = {38},
  number = {1},
  pages = {28},
  issn = {2210-5441},
  doi = {10.1007/s13347-025-00857-w},
  urldate = {2025-09-16},
  langid = {english}
}

@article{liuArtificialIntelligencePerceived2024,
  title = {Artificial Intelligence and Perceived Effort in Relationship Maintenance: {{Effects}} on Relationship Satisfaction and Uncertainty},
  shorttitle = {Artificial Intelligence and Perceived Effort in Relationship Maintenance},
  author = {Liu, Bingjie and Kang, Jin and Wei, Lewen},
  year = {2024},
  month = may,
  journal = {Journal of Social and Personal Relationships},
  volume = {41},
  number = {5},
  pages = {1232--1252},
  publisher = {SAGE Publications Ltd},
  issn = {0265-4075},
  doi = {10.1177/02654075231189899},
  urldate = {2025-09-16},
  langid = {english}
}

@article{romeroWritingWrongsPromoting2008,
  title = {Writing Wrongs: {{Promoting}} Forgiveness through Expressive Writing},
  shorttitle = {Writing Wrongs},
  author = {Romero, Catherine},
  year = {2008},
  month = aug,
  journal = {Journal of Social and Personal Relationships},
  volume = {25},
  number = {4},
  pages = {625--642},
  publisher = {SAGE Publications Ltd},
  issn = {0265-4075},
  doi = {10.1177/0265407508093788},
  urldate = {2025-09-16},
  langid = {english}
}

@inproceedings{cerejoAnticipationToolDesigning2024,
  title = {Anticipation as a {{Tool}} for {{Designing}} the {{Future}}},
  booktitle = {Advances in {{Design}} and {{Digital Communication IV}}},
  author = {Cerejo, Joana and Carvalhais, Miguel},
  editor = {Martins, Nuno and Brand{\~a}o, Daniel},
  year = {2024},
  pages = {37--52},
  publisher = {Springer Nature Switzerland},
  address = {Cham},
  doi = {10.1007/978-3-031-47281-7_4},
  isbn = {978-3-031-47281-7},
  langid = {english}
}

@article{zamenopoulosAnticipatoryViewDesign2007,
  title = {Towards an Anticipatory View of Design},
  author = {Zamenopoulos, Theodore and Alexiou, Katerina},
  year = {2007},
  month = jul,
  journal = {Design Studies},
  volume = {28},
  number = {4},
  pages = {411--436},
  issn = {0142-694X},
  doi = {10.1016/j.destud.2007.04.001},
  urldate = {2025-09-16}
}

\end{document}